\documentclass[aip,rsi,reprint,graphicx, amsmath,amssymb,amsfonts]{revtex4-1} 
\usepackage{subfigure}
\usepackage{psfrag}
\usepackage[dvips]{graphicx}
\usepackage{color}
\usepackage{aas_macros}
\usepackage{multirow}
\usepackage[normalem]{ulem}

\begin{document}


\title{Systematic effects from an ambient-temperature, continuously-rotating half-wave plate} 



\author{T.~Essinger-Hileman}
\affiliation{Department of Physics, Princeton University, Princeton, New Jersey, 08544 USA}
\affiliation{Department of Physics and Astronomy, The Johns Hopkins University, Baltimore, Maryland, 21218 USA}

\author{A.~Kusaka}
\email{akusaka@lbl.gov}
\affiliation{Department of Physics, Princeton University, Princeton, New Jersey, 08544 USA}
\affiliation{Physics Division, Lawrence Berkeley National Laboratory, Berkeley, California, 94720 USA}

\author{J.~W.~Appel}
\affiliation{Department of Physics, Princeton University, Princeton, New Jersey, 08544 USA}
\affiliation{Department of Physics and Astronomy, The Johns Hopkins University, Baltimore, Maryland, 21218 USA}
\author{S.~K.~Choi}
\author{K.~Crowley}
\affiliation{Department of Physics, Princeton University, Princeton, New Jersey, 08544 USA}
%
%
\author{S.~P.~Ho}
\affiliation{Department of Physics, Princeton University, Princeton, New Jersey, 08544 USA}
\author{N.~Jarosik}
\affiliation{Department of Physics, Princeton University, Princeton, New Jersey, 08544 USA}
%
%
\author{L.~A.~Page}
\affiliation{Department of Physics, Princeton University, Princeton, New Jersey, 08544 USA}
\author{L.~P.~Parker}
\affiliation{Department of Physics, Princeton University, Princeton, New Jersey, 08544 USA}
\affiliation{Department of Physics and Astronomy, The Johns Hopkins University, Baltimore, Maryland, 21218 USA}
\author{S.~Raghunathan}
\affiliation{Department of Astronomy, Universidad de Chile, Santiago, Chile}
\affiliation{School of Physics, University of Melbourne, Parkville, VIC 3010, Australia}
%
%
\author{S.~M.~Simon}
\author{S.~T.~Staggs}
\author{K.~Visnjic}
\affiliation{Department of Physics, Princeton University, Princeton, New Jersey, 08544 USA}




\date{\today}

\begin{abstract}
We present an evaluation of systematic effects associated with a continuously-rotating, ambient-temperature half-wave plate (HWP) based on two seasons of data from the Atacama B-Mode Search (ABS) experiment located in the Atacama Desert of Chile. The ABS experiment is a microwave telescope sensitive at 145~GHz.  Here we present our in-field evaluation of celestial (CMB plus galactic foreground) temperature-to-polarization leakage. We decompose the leakage into scalar, dipole, and quadrupole leakage terms. We report a scalar leakage of $\sim 0.01\%$, consistent with model expectations and an order of magnitude smaller than other CMB experiments have reported. No significant dipole or quadrupole terms are detected; we constrain each to be  $< 0.07$\% (95\% confidence), limited by statistical uncertainty in our measurement.
 Dipole and quadrupole leakage at this level lead to systematic error on $r \lesssim 0.01$ before any mitigation due to scan cross-linking or boresight rotation.
 The measured scalar leakage and the theoretical level of 
 dipole and quadrupole leakage produce systematic error 
 of $r < 0.001$ for the ABS survey and focal-plane layout before 
 any data correction such as so-called deprojection.
 This demonstrates that ABS achieves significant beam systematic error mitigation from its HWP and shows the promise of continuously-rotating HWPs for future experiments.

\end{abstract}


\maketitle 


\section{Introduction}
Precise measurements of the Cosmic Microwave Background (CMB) polarization provide a unique window into the physics of the very early universe, where quantum-gravitational effects are expected to play an important role. A primordial gravitational-wave background (GWB) would leave a unique odd-parity ``B-mode'' pattern in the CMB polarization.\cite{1997PhRvL..78.2054S,1997PhRvL..78.2058K} Many models of inflation predict an observable GWB.\cite{2015arXiv151006042K} Its amplitude, as imprinted in the CMB polarization, is a direct measure of the energy scale of inflation. A detection of gravitational-wave-induced B-mode polarization in the CMB would provide compelling evidence for inflation and a rare glimpse into physics at ultra-high energies. The level of B-modes is parametrized by the tensor-to-scalar ratio, $r$, which is currently constrained to be $< 0.07$ (95\% confidence).\cite{2015arXiv151009217A}

CMB polarization experiments face a daunting task as the level of the B-mode polarization is well below the level of unpolarized foregrounds. This makes systematic errors due to temperature-to-polarization leakage particularly detrimental. Polarization modulators offer a means of separating the polarized signal of interest from these unpolarized foregrounds. Many polarization modulation schemes exist \cite{1946RScI...17..268D, 2003ApJS..145..413J, 2004ApJ...610..625F, 2005ApJS..159....1B, 2012ApJ...760..145Q, 2010A&A...520A...4B, 2006PhDT........33S, 2003PhRvD..68d2002O, 2005ApJ...624...10L, CBI_instrument_2002, 2009ApJ...694.1664C, 2013ApJ...765...64M}, and a rapidly-rotating half-wave plate (HWP)\cite{1988PASP..100.1158J, 1991PASP..103.1193P, 1991ApJ...370..257L, 2007ApJ...665...42J, JohnRuhl2008, 2014RScI...85b4501K, 2010SPIE.7741E..37R, 2015JLTP..tmp...76R} is one of the most promising. One of the key advantages of HWP modulation is that it allows single polarization-sensitive detectors to act as complete $Q/U$ polarimeters. Without modulation, experiments gain polarization sensitivity by differencing the output of pairs of detectors with sensitivity to orthogonal linear polarizations; however, pair differencing can cause significant temperature-to-polarization leakage if, for example,  the pair of detectors has mismatched beams.\cite{2008PhRvD..77h3003S} The BICEP2 and Keck Array teams estimate that their deprojection analysis technique reduces $I \rightarrow Q/U$ leakage by a factor of ten or more in their maps, to the $r=0.003$ level.~\cite{2015ApJ...811..126A} 
Using a rapidly-rotating HWP eliminates the need for beam differencing and reduces the requirements on analysis techniques for removing any residual leakage contamination.

The Atacama B-Mode Search (ABS) experiment consists of 240 feedhorn-coupled bolometric polarimeters observing at 145 GHz with a rapidly-rotating, ambient-temperature HWP at the entrance aperture near a stop.\cite{2009AIPC.1185..494E, JohnAppelPrincetonThesis, EssingerHilemanPrincetonThesis, LucasParkerPrincetonThesis, KaterinaVisnjicPrincetonThesis} Having the modulator as the first optical element in the system allows clear separation of instrumental polarization from celestial polarization; however, we note that the modeling presented below does not require the HWP to be at a stop, allowing straightforward application to systems such as those of POLARBEAR~\cite{2014JLTP..176..719S} and the Atacama Cosmology Telescope.~\cite{2013AAS...22110504N, 2015arXiv151002809H} The HWP is made of single-crystal, $\alpha$-cut sapphire 330 mm in diameter and 3.15 mm thick. It is designed to work at 145~GHz. Sapphire has ordinary and extraordinary indices of refraction of $3.068\pm0.003$ and $3.402\pm0.003$, respectively.~\cite{1994IJIMW..15..339P} It is anti-reflection (AR) coated with 305 $\mu$m of Rogers RT/Duroid 6002,\cite{KaterinaVisnjicPrincetonThesis} a glass-reinforced PTFE laminate with a refractive index of $1.715\pm0.012$.\footnote{https://www.rogerscorp.com/acs/products/34/RT-duroid-6002-Laminates.aspx} An air-bearing system allows the HWP to rotate at a stable frequency of 2.55 Hz. Porous graphite pads\footnote{ {N}ewWay Air Bearings, 50 McDonald Blvd, Aston, PA 19014 USA} are placed around an aluminum rotor at three points on its circumference. Compressed air is forced through the graphite to float the rotor with almost no friction. An incremental encoder disc with an index to mark the zero point is used to read out the HWP angle with $2.4^{\prime}$ resolution. 

The detectors for ABS were fabricated in two separate batches, which we label A and B, with half of the detectors in each batch. Due to an unexpected change in the microstrip dielectric constant between fabrications, batch B has a bandpass shifted up by $\sim 12$ GHz. The HWP is optimized for the bandpass of batch A, which carries approximately 90\% of the statistical weight in the maps. The two batches of detectors are in separate halves of the focal plane, but both have detectors across the full range of radii from the center, which is relevant for comparison to the model.

\begin{figure}[htbp]
	\centering
	\includegraphics[width=0.35\textwidth, clip=true, trim=0cm 3cm 0cm 3.5cm]{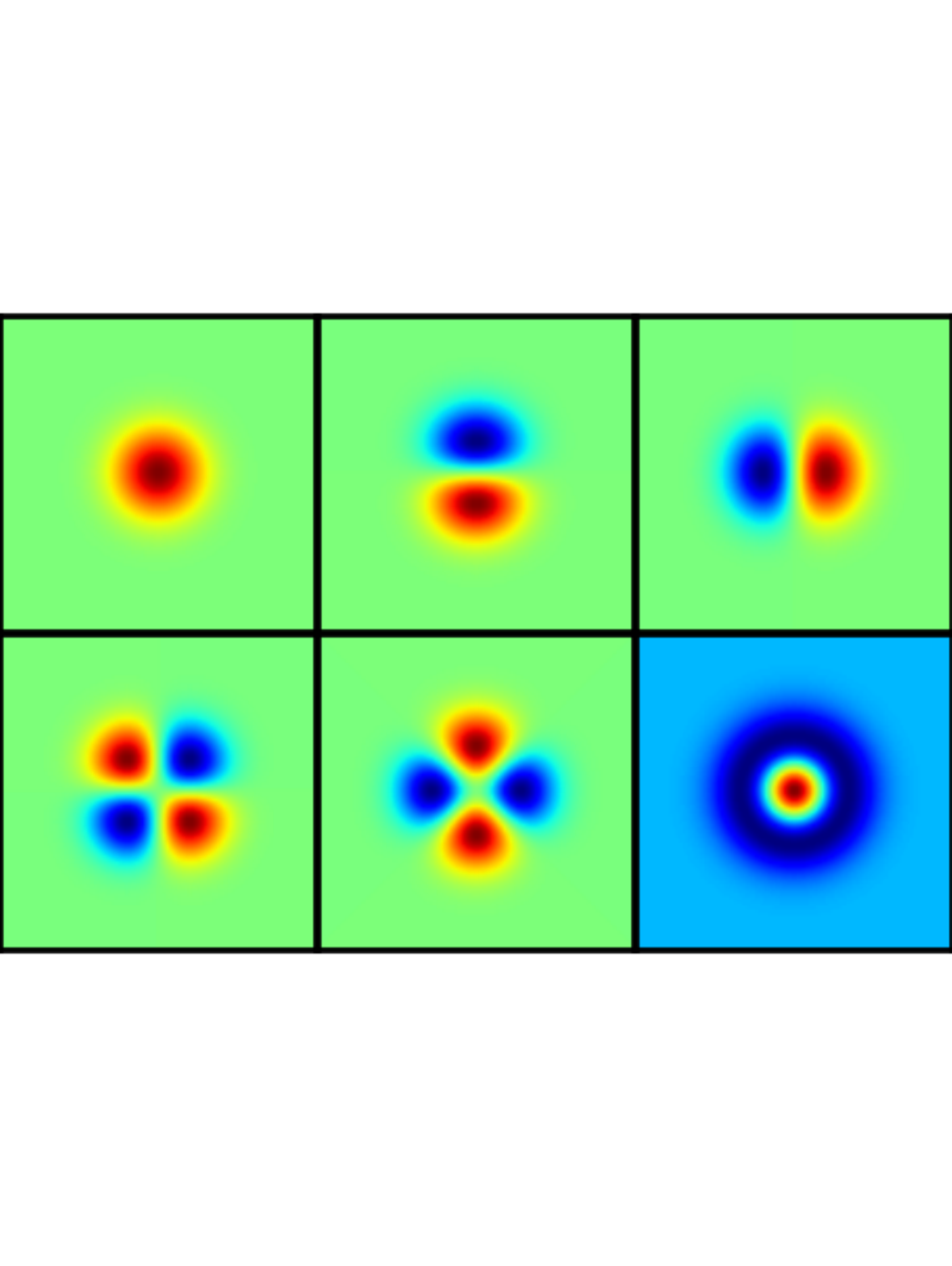}
	\caption{Monopole (``scalar'', top left), dipole (top middle and right), quadrupole (bottom left and middle), and differential width (bottom right) functions used to expand the modulated beam.
        In these maps, red and blue envelopes correspond to positive and negative leakages, respectively.
        The HWP does not induce differential width beam distortions. The functions are defined in terms of Gauss-Hermite functions, as in Equation \ref{eqn:gauss_hermite_functions}. }
	\label{fig:gauss_hermite_expansion}
\end{figure}

The ABS HWP allows for separation of unpolarized atmospheric fluctuations, unpolarized ground pickup, and instrumental polarization from celestial polarization. In a companion paper,\cite{2014RScI...85b4501K} we demonstrated the ability of ABS to reject atmospheric fluctuations at better than 30 dB at 2 mHz. Here we present our in-field evaluation of celestial temperature-to-polarization leakage based on two seasons of observations. We break the leakage down into scalar, dipole, and quadrupole terms\cite{2008PhRvD..77h3003S} (see Figure~\ref{fig:gauss_hermite_expansion}) and investigate their effects on the power spectra from ABS. This parametrization is similar to the differential gain, pointing, and ellipticity in beam subtraction experiments, which we compare to in Section~\ref{sec:higher}. In Section~\ref{sec:beam_decomp} we consider systematics associated with the beam profile and model expectations of their levels and functional forms. We characterize the scalar leakage in ABS in Section~\ref{sec:scalar} and higher-order terms in Section~\ref{sec:higher}, along with their impact on constraints on  $r$. We conclude in Section~\ref{sec:conclusion}.

\section{Description of systematic effect associated with the beam profile}
\label{sec:beam_decomp}
The ABS data are demodulated in order to separate polarized from unpolarized emission.\cite{2014RScI...85b4501K} The HWP modulates incoming linear polarization at four times its rotation frequency $f_m$. The resulting data depend on the input Stokes parameters $(I,Q,U)$ and HWP angle, $\chi$, as
\begin{equation}
 \label{equ:d_m_tod}
d_{m} = I + \varepsilon \mbox{Re} \left[ (Q+i U) m(\chi) \right] + A(\chi) + \mathcal{N},
\end{equation}

\noindent where $m(\chi) \equiv \exp{\left(-i 4 \chi \right)}$ is the modulation function, $A(\chi)$ is a HWP synchronous signal, $\varepsilon$ is the polarization modulation efficiency, and $\mathcal{N}$ is a noise term. We describe $A(\chi)$ in terms of its two dominant components
\begin{equation}
A(\chi) = A_{0} (\chi) + \lambda(\chi) I,
 \label{equ:a_of_chi}
\end{equation}

\noindent where $A_{0} (\chi)$ is independent of sky intensity, and the second term corresponds to conversion of unpolarized sky signal to modulated polarized light. Both components can be expanded as Fourier series in $\chi$, with the largest component in the $\sin 2 \chi$ and $\cos 2 \chi$ terms, corresponding to modulation in the detector timestreams at $2 f_{m}$. Here we focus on the smaller $4 f_{m}$ components, which cause leakage into the polarized signal of interest. (The $2f_{m}$ components can be used to assess data quality.~\cite{2015arXiv151104760S}) The $4 f_{m}$ components arise from reflection-induced polarization being rotated, and thus modulated, by the HWP. Thus temperature-to-polarization leakage systematics increase with increasing angle of incidence, going to zero at normal incidence.

By multiplying the timestream by the complex conjugate of the modulation function and lowpass filtering below the modulation frequency, we create a complex-valued demodulated timestream:
\begin{equation}
\label{demodulated_tod}
\begin{split}
d_{\bar{d}} = \frac{1}{2} \left( \varepsilon_{Q} Q + \lambda_{Q} I + \xi_{QU} U + A_{0}^{Q} \right) +  \mathcal{N}^{Re} + \\ \frac{i}{2} \left( \varepsilon_{U} U + \lambda_{U} I + \xi_{UQ} Q + A_{0}^{U} \right) + i \mathcal{N}^{Im} \;.
\end{split}
\end{equation}

\noindent The real and imaginary parts of $d_{\bar{d}}$ are equivalent to $Q$ and $U$. Small $Q/U$ leakage terms have been added to this equation, denoted by $\xi_{QU}$ and $\xi_{UQ}$, which occur due to the same mechanism as temperature-to-polarization leakage, namely reflection-induced polarization. The total-power timestream is constructed separately by removing $A(\chi)$ and/or lowpass filtering the data. The $A(\chi)$ removal is done by binning all data in a certain time span (typically $\sim$1 hour) versus HWP angle $\chi$ and then subtracting this waveform from the data. 
We note that Eq.~(\ref{demodulated_tod}) ignores a very
small effect where $I$ signal placed at around $4 f_m$ frequency by scan modulation
can remain as a residual temperature-to-polarization leakage. In contrast to the
$\lambda_Q I$ and $\lambda_U I$ terms in Eq.~(\ref{demodulated_tod}),
this effect couples two different angular scales; it is a leakage from
intensity at small angular scales ($\sim 1\,\mathrm{arcmin}$ for the ABS
scan speed) to degree-scale polarization. The effect is
suppressed because of the beam size of the instrument and the
fact that fine-resolution, or high-$\ell$, signal is small for CMB.
We also note that this
effect is not intrinsic to the instrument.  If necessary, a map-making
process can completely eliminate this systematic effect;
 a simple example of such a method is to take the difference between two orthogonal
detectors in addition to the demodulation.

\begin{figure*}[htbp]
	\centering
	\subfigure[ \label{fig:model_IP_mono}]{\includegraphics[width=0.4\textwidth, clip=true, trim=0cm 0cm 0cm 1cm]{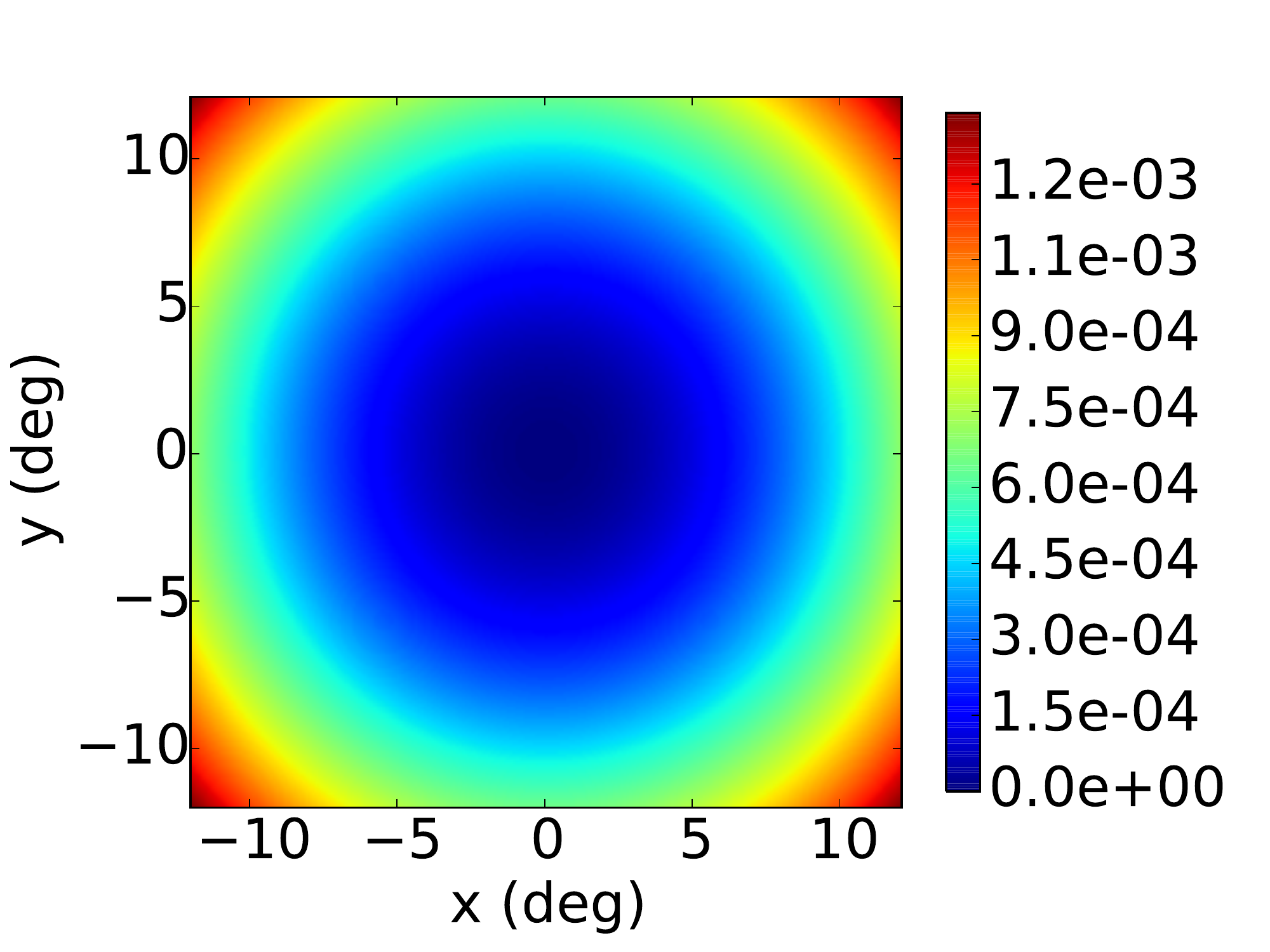}}
	\hspace{0.03\textwidth}
	\subfigure[\label{fig:model_residual}]{\includegraphics[width=0.4\textwidth, clip=true, trim=0cm 0cm 0cm 1cm]{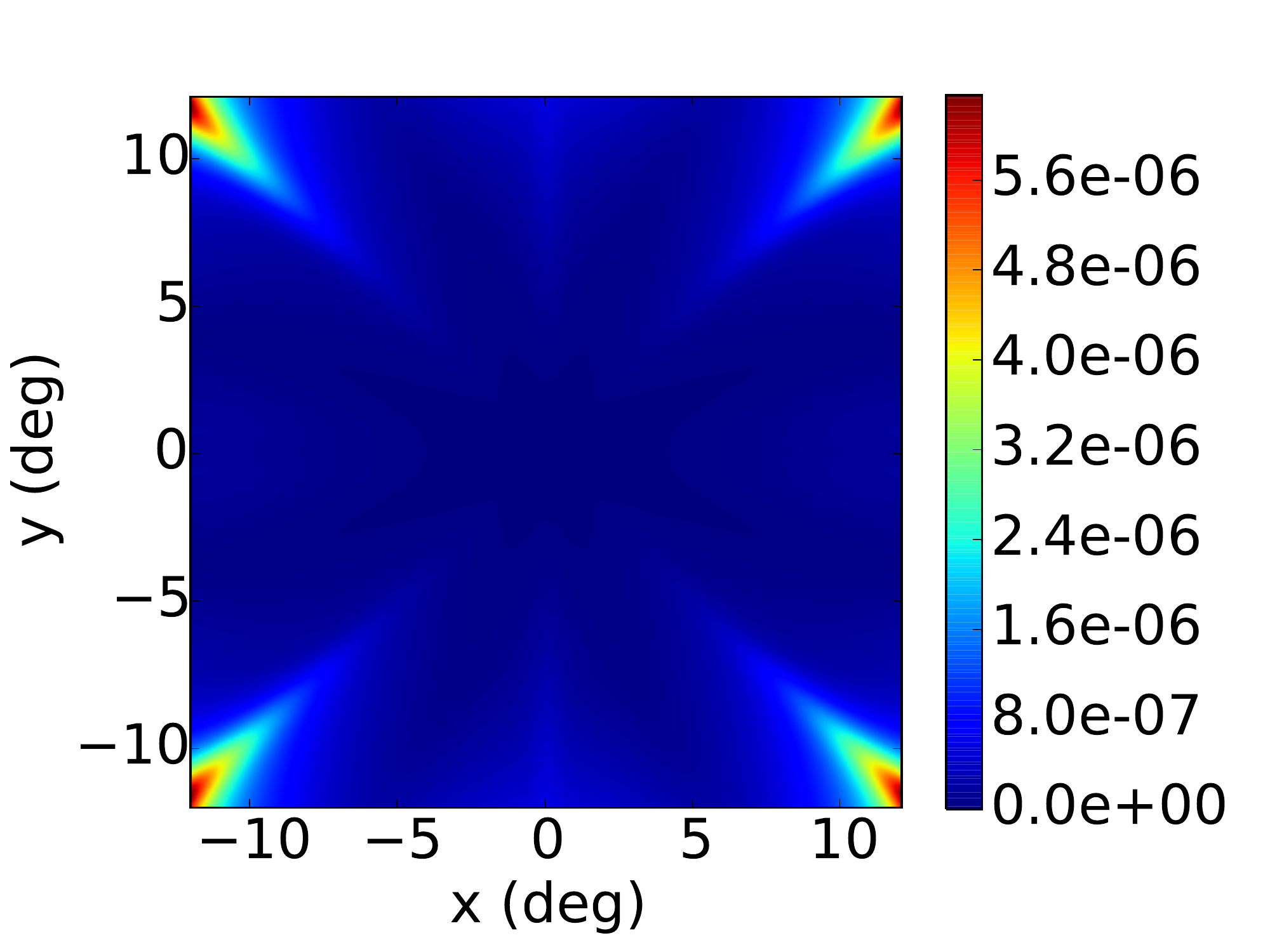}}
	\caption{ %
		(a) Total leakage $\Lambda_{P}$ and (b) residual of Gauss-Hermite fit up to second order (monopole, dipole, and quadrupole terms) versus offset angle of a detector line-of-sight from instrument boresight for a $32^{\prime}$ FWHM Gaussian beam as estimated by the transfer-matrix model\cite{2013ApOpt..52..212E} at 145 GHz. The colorbar indicates the integrated power normalized to that of an unpolarized input Gaussian beam. The majority of the leakage beam is captured in the monopole, dipole, and quadrupole terms.}
	\label{fig:model_leak_residual}
\end{figure*}

%
We project the focal plane on the sky, and 
specify a position on the focal plane as well as the angular dependence
of the instrument's response using a spherical coordinate system,
$\left(\theta,\phi\right)$.
Here, $\theta$ is measured relative to the line of sight (LOS) corresponding
to the center of the focal plane, and $\phi$ describes rotation about
this LOS.
%
The LOS to a detector $k$ is described by $\left(\theta_{k0},\phi_{k0}\right)$. The focusing optics of the instrument define the beam of each detector, $B_{k} (\theta_{k}, \phi_{k})$, where $\left(\theta_{k},\phi_{k}\right)$ are defined relative to the detector LOS at $\left(\theta_{k0},\phi_{k0}\right)$. For this work we take the $B_{k}$ to be azimuthally symmetric Gaussians each with a full width at half maximum (FWHM) of $32^{\prime}$, a reasonable approximation for ABS. Since for ABS the HWP is the first element in the optical chain, the spread of rays impinging on the HWP for detector $k$ is described by that detector's beam function in the global instrument coordinates. The HWP modifies incoming polarization differently depending upon the angle of rays going through it and those rays' polarization state. These two effects combine to make $\varepsilon$ different functions for $Q$ and $U$ Stokes parameters, $\varepsilon_{Qk} (\theta_{k}, \phi_{k})$ and $\varepsilon_{Uk} (\theta_{k}, \phi_{k})$. The leakage beams, $\lambda_{Qk}$ and $\lambda_{Uk}$, similarly have angular dependence. Thus, Equation \ref{demodulated_tod}  is the result of an integral over $\theta_{k}$ and $\phi_{k}$.

To quantify systematic errors induced by the HWP, we seek to characterize (1) the beam-averaged magnitudes of $\lambda_{Q}$ and $\lambda_{U}$, denoted by $\Lambda_{Q}$ and $\Lambda_{U}$, which cause direct temperature-to-polarization (scalar) leakage, (2) the higher multipole terms in $\lambda_{Q}$ and $\lambda_{U}$; and, (3) how $\lambda_{Q}$ and $\lambda_{U}$ vary for detectors at different $\theta_{k0}$ and $\phi_{k0}$ across the focal plane. Another interesting property we can model is the small deviation of the polarized beams, $\varepsilon_{Qk} \left(\theta_{k}, \phi_{k} \right)$ and $\varepsilon_{Uk} \left(\theta_{k}, \phi_{k} \right)$, from Gaussians, which we will consider in a future paper. We define the total leakage as 
\begin{equation}
\Lambda_{P} \equiv \sqrt{\Lambda_{Q}^{2} + \Lambda_{U}^{2}}.
\label{eqn:total_leak_def}
\end{equation}


\begin{figure*}[htbp]
\centering
 \subfigure[I-to-Q leakage monopole, $a_{m0}$, for $\lambda_{Q}$, which is equivalent to $\Lambda_{Q}$. This is also the shape of $g_{I}^{4C}$ in Equation \ref{eqn:mod_mueller_2}.\label{fig:IQ_mono_leak}]{\includegraphics[width=0.4\textwidth, clip=true, trim=0cm 0cm 0cm 0cm]{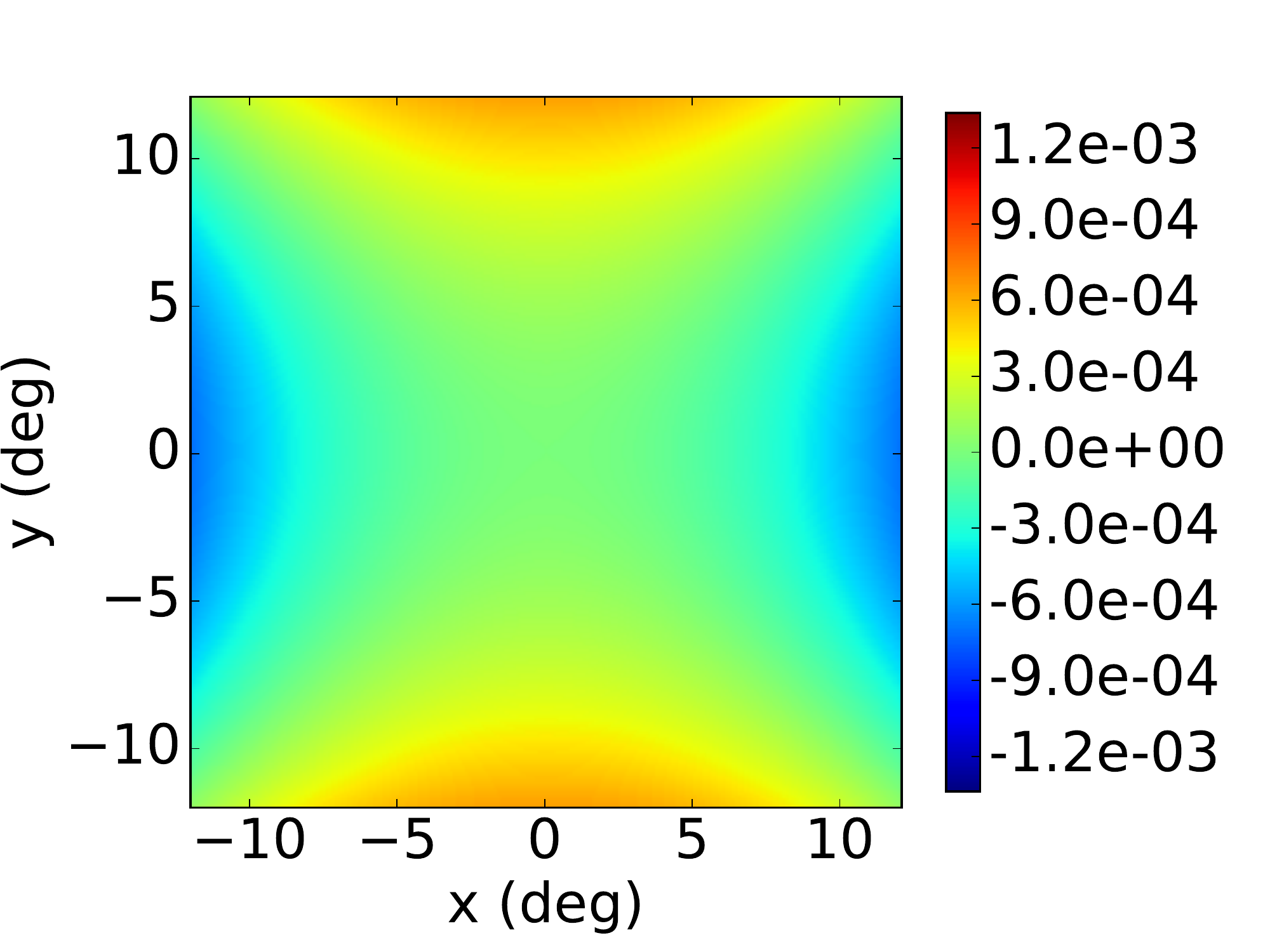}}
 \hspace{0.03\textwidth}
 \subfigure[I-to-U leakage monopole, $a_{m0}$, for $\lambda_{U}$, which is equivalent to $\Lambda_{U}$. This is also the shape of $g_{I}^{4S}$ in Equation \ref{eqn:mod_mueller_2}.\label{fig:IU_mono_leak}]{\includegraphics[width=0.4\textwidth, clip=true, trim=0cm 0cm 0cm 0cm]{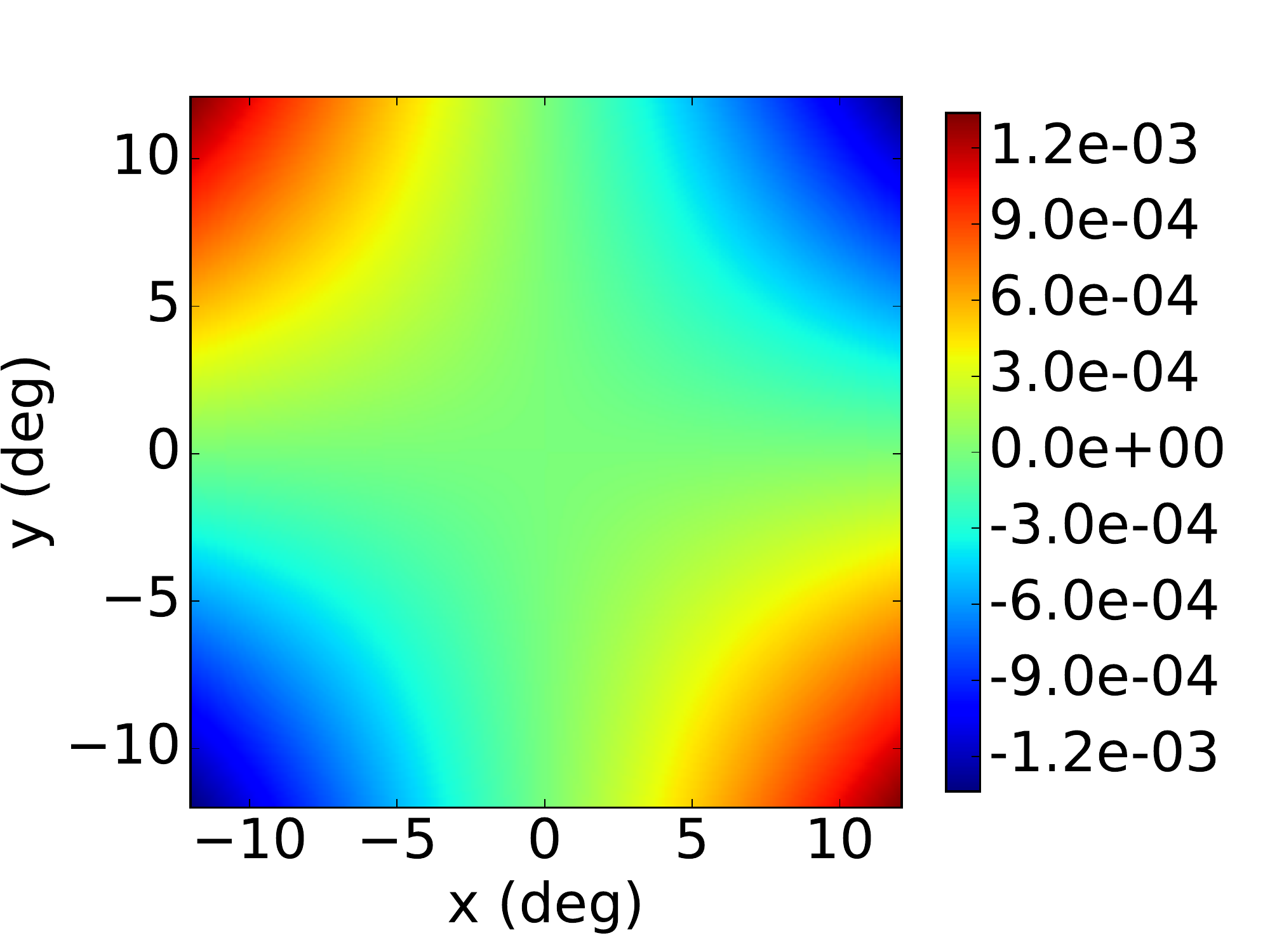}}
 \subfigure[I-to-Q dipole leakage. The color scale denotes total dipole leakage $\sqrt{a_{d1}^{2} + a_{d2}^{2}}$ for $\lambda_{Q}$. The arrows show the direction of the dipole, pointing toward the positive lobe of the dipole. \label{fig:IQ_dip_leak}]{\includegraphics[width=0.4\textwidth, clip=true, trim=0cm 0cm 0cm 1cm]{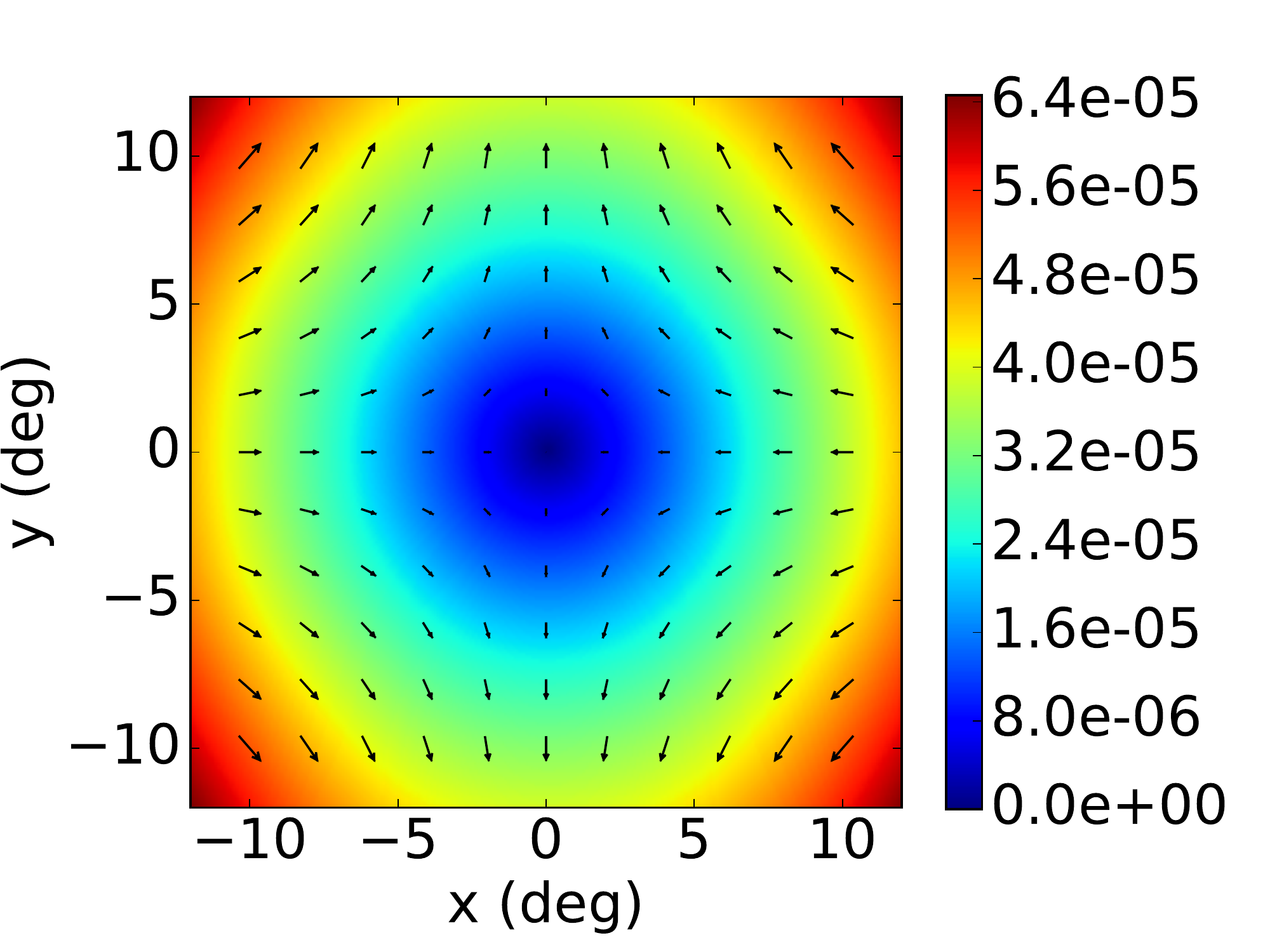}}
   \hspace{0.03\textwidth}
 \subfigure[I-to-U dipole term. The color scale denotes total dipole leakage $\sqrt{a_{d1}^{2} + a_{d2}^{2}}$ for $\lambda_{U}$. The arrows show the direction of the dipole, pointing toward the positive lobe of the dipole.\label{fig:IU_dip_leak}]{\includegraphics[width=0.4\textwidth, clip=true, trim=0cm 0cm 0cm 1cm]{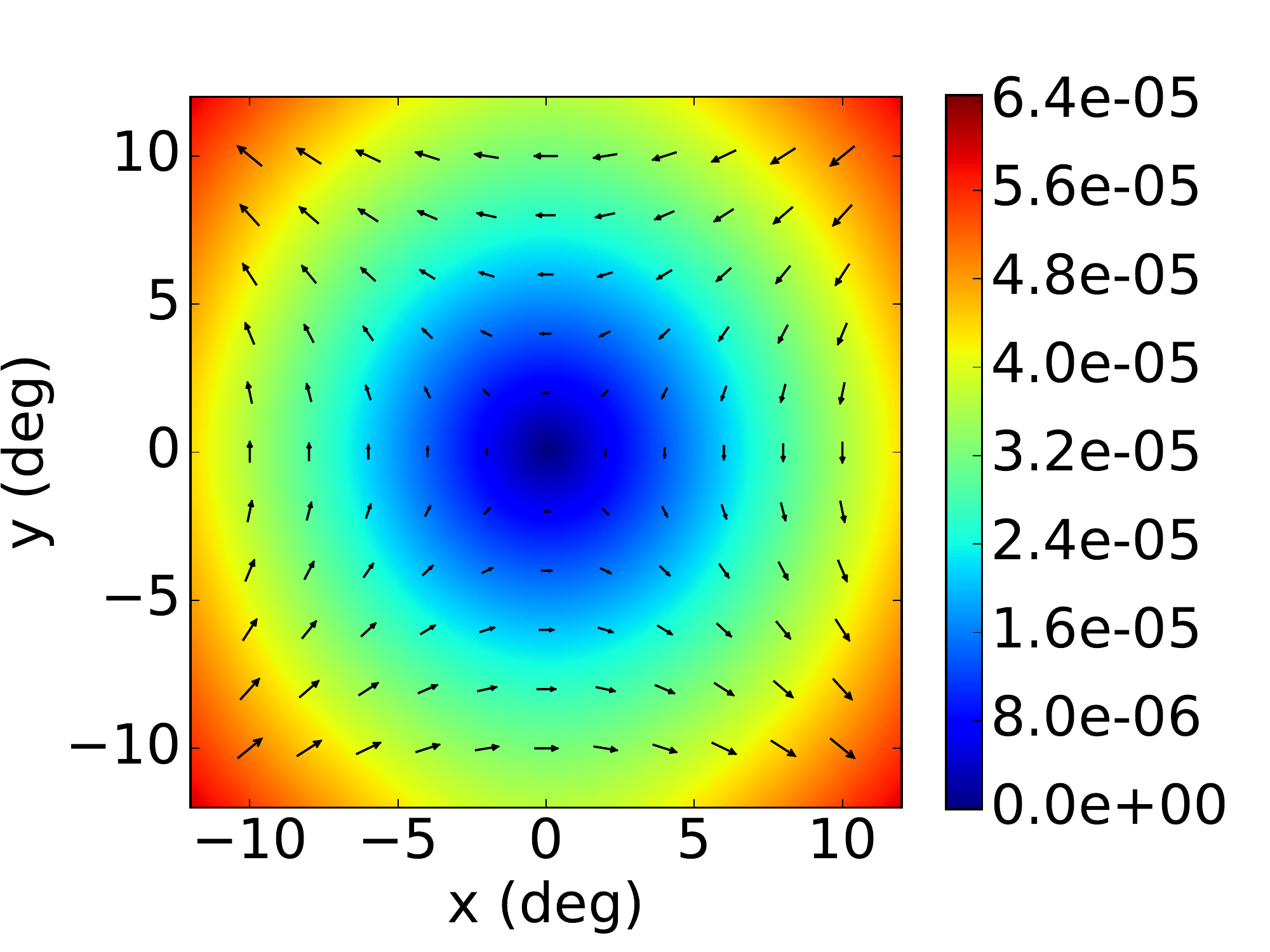}}
 \subfigure[I-to-Q quadrupole term. The color scale denotes total quadrupole leakage $\sqrt{a_{q1}^{2} + a_{q2}^{2}}$ for $\lambda_{Q}$. The solid (dashed) line shows the direction of the positive (negative) lobes of the quadrupole.\label{fig:IQ_quad_leak}]{\includegraphics[width=0.4\textwidth, clip=true, trim=0cm 0cm 0cm 1cm]{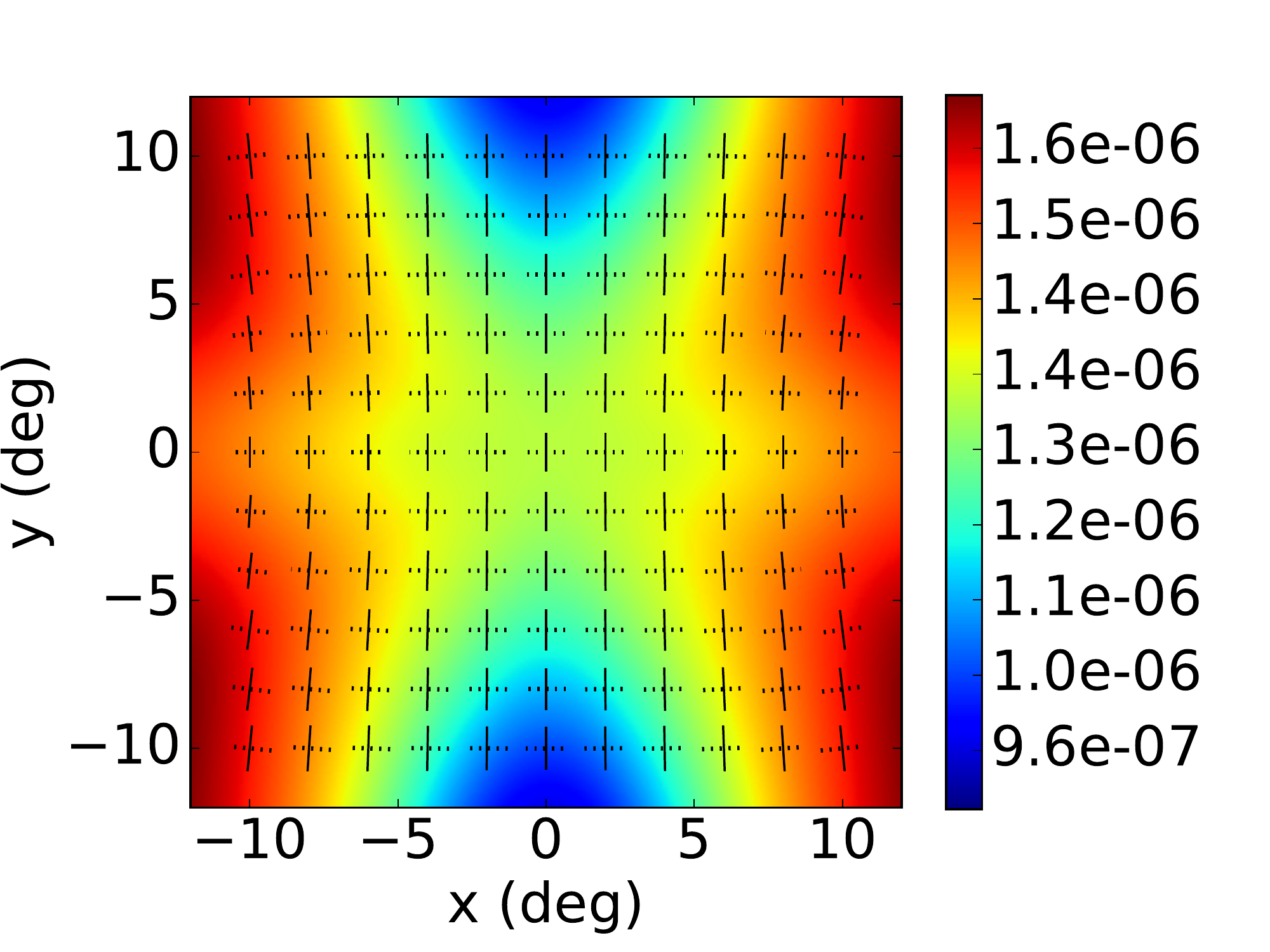}}
   \hspace{0.03\textwidth}
 \subfigure[I-to-U quadrupole term. The color scale denotes total quadrupole leakage $\sqrt{a_{q1}^{2} + a_{q2}^{2}}$ for $\lambda_{U}$. The solid (dashed) line shows the direction of the positive (negative) lobes of the quadrupole.\label{fig:IU_quad_leak}]{\includegraphics[width=0.4\textwidth, clip=true, trim=0cm 0cm 0cm 1cm]{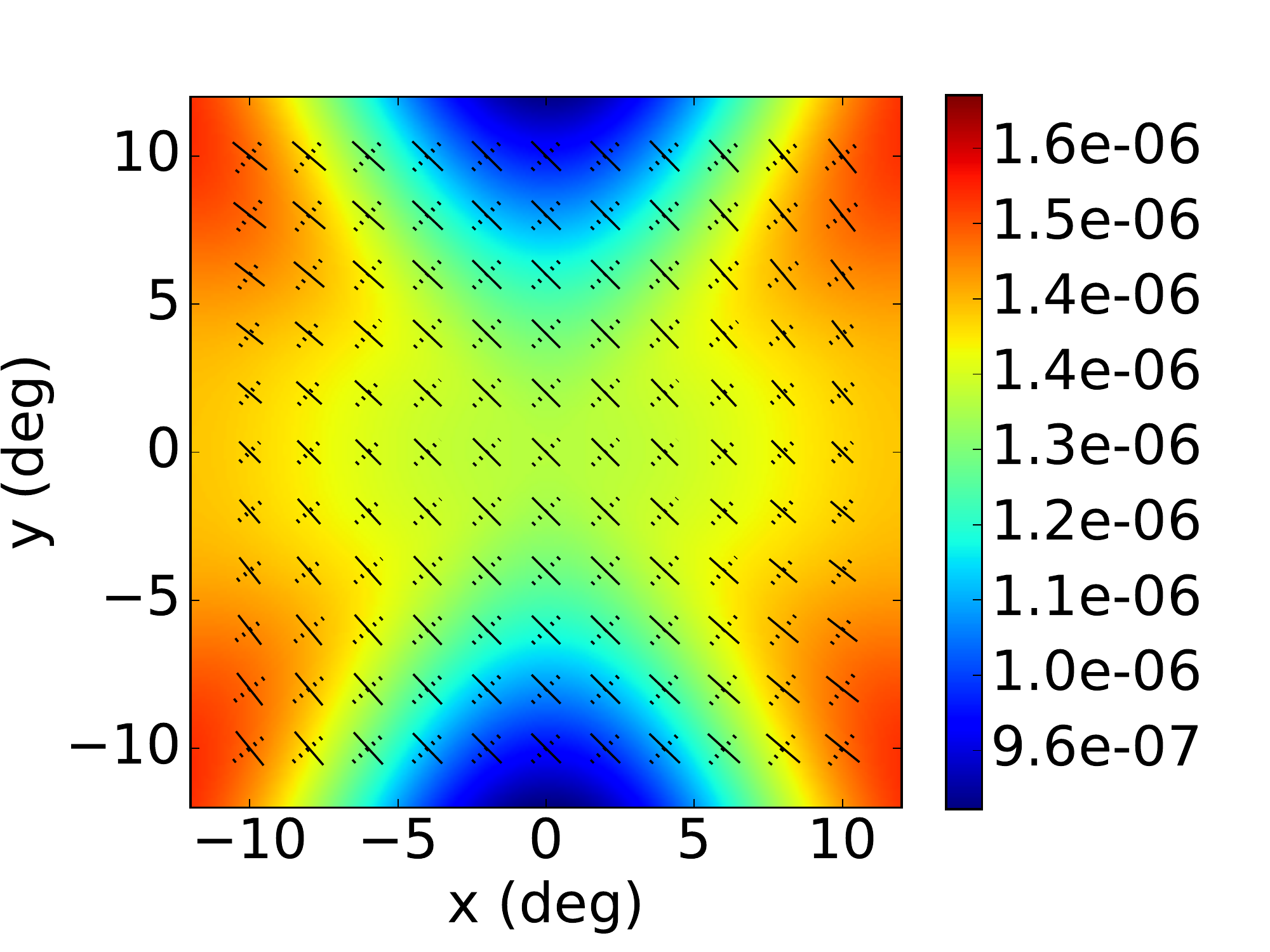}}
 \caption{ %
 Modeled monopole, dipole, and quadrupole leakage terms versus offset angle from boresight for a $32^{\prime}$ FWHM Gaussian beam as estimated by the transfer-matrix model\cite{2013ApOpt..52..212E} at 145 GHz for the ABS HWP. For definitions of the $a$ coefficients, see Equation \ref{eqn:a_coeffs}. The colorbars indicate the integrated power in each component normalized to the unpolarized integrated input power.
 }
\label{fig:model_outputs}
\end{figure*}

\subsection*{Modeling beam systematic effects}
To estimate the HWP beam systematic effects described above, a model of the HWP has been developed based upon the $4 \times 4$ transfer-matrix method.\cite{2013ApOpt..52..212E} We can write the system response for ABS in matrix form, similar to a Mueller matrix, where the outputs of the data-reduction pipeline are total-power $\hat{I}$ and the real and imaginary parts of the demodulated timestream, Equation \ref{demodulated_tod}, which correspond to $\cos 4 \chi$ and $\sin 4 \chi$ components. The input is the Stokes vector, $\vec{S} = (I,Q,U,V)$. This means the system is described by an angle-dependent $4 \times 3$ matrix, $\boldsymbol{M}_\mathrm{total}$, that maps the Stokes vector on the sky to the outputs of the data-reduction pipeline:
\begin{equation}
  \left(
   \begin{array}{c}
    \hat{I} \\
    \hat{Q} \\
	\hat{U}
   \end{array}
  \right)
   = 
 \int  \left( \boldsymbol{M}_\mathrm{total} \cdot \vec{S}  \right) d \Omega .
\label{eqn:abs_system_response}
\end{equation}

\noindent Here $\boldsymbol{M}_\mathrm{total}$ can be expanded in terms of the functions defined earlier,
\begin{eqnarray}
\label{total_mueller}
\boldsymbol{M}_\mathrm{total}
  & \equiv & \hspace{48pt}
  \left(
  \begin{array}{cccc}
  B_{k} & 0 & 0 & 0 \\
  \lambda_{Qk} & \varepsilon_{Qk} & \xi_{QUk}  & \zeta_{Qk} \\
  \lambda_{Uk} & \xi_{UQk} & \varepsilon_{Uk}  & \zeta_{Uk} \\
  \end{array}
  \right)  \\ \label{eqn:mod_mueller_2}
  & = & B_{k} \left(\theta_{k}, \phi_{k} \right)
  \left(
   \begin{array}{cccc}
    1 & 0 & 0 & 0 \\
     g_{I}^{4C} & g_{Q}^{4C} & g_{U}^{4C}  & g_{V}^{4C} \\
     g_{I}^{4S} & g_{Q}^{4S} & g_{U}^{4S} & g_{V}^{4S} \\
   \end{array} 
  \right) . 
\end{eqnarray}

\noindent The nonzero terms of $M_\mathrm{total}$ depend on $\theta$ and $\phi$. The six angular response functions $g^{4C,4S}_{I,Q,U} (\theta, \phi)$ are calculated with the transfer-matrix model. The terms $g_{V}^{4C,4S}$ are expected to be negligible, as is celestial circular polarization, so we neglect  $\zeta_{Qk}$ and  $\zeta_{Qk}$ here. The model results are calculated at 145 GHz at the center of the ABS band. This frequency is relevant for systematic effects such as internal reflection caused by imperfect anti-reflection coating for off-axis incident rays. No band averaging is performed. The model describes how the HWP would couple infinite plane waves with incoming Stokes $I$, $Q$, and $U$ parameters to the real and imaginary parts of the demodulated timestream in the absence of the focusing optics. 
 
Multiplying the $g$ functions by the beam yields $\lambda_{Qk} \left( \theta_{k}, \phi_{k} \right)$ and $\lambda_{Uk} \left( \theta_{k}, \phi_{k} \right)$ and the other terms in Equation \ref{demodulated_tod}. This amounts to decomposing the Gaussian beam at the HWP into a superposition of infinite plane waves. This approach ignores edge effects due to truncation of the beam by the telescope aperture, which is a valid approximation for ABS because the aperture is many, $\simeq 100$, wavelengths across.


\subsection*{Modeled leakage levels}
We wish to characterize the impact of HWP beam systematics in a way that highlights the effects on final data quality. To this end, we perform an expansion of the beam into monopole (scalar), dipole, and quadrupole terms.\cite{2008PhRvD..77h3003S, 2015arXiv150200596B} The expansion functions are shown in Figure \ref{fig:gauss_hermite_expansion}. We focus here on the temperature-to-polarization leakage beams, $\lambda_{kQ},\lambda_{kU}$, which can be expanded into Gauss-Hermite functions as
\begin{equation}
\lambda_{k \{Q,U\}} (\theta_{k}, \phi_{k}) = \sum_{j=0}^{\infty} \sum_{i=0}^{\infty} s_{kij} f_{ij} (\theta_{k}, \phi_{k}) . \\
\label{eqn:lambda_expansion}
\end{equation}

\noindent Here $s_{kij}$ are the fit coefficients and the normalized basis functions $f_{ij} (\theta, \phi)$ are 
\begin{equation}
 \begin{split}
f_{ij} (\theta_{k}, \phi_{k}) = & \left( \frac{\exp{\left[- \theta_k^2/(2 \sigma^{2})\right]}}{\sqrt{2^{i+j} i! j! \pi \sigma^{2}}} \right)
  \\
  & \times H_{i} \left( \frac{\theta_{k}\cos\phi_{k}}{\sigma} \right)  H_{j} \left( \frac{\theta_{k}\sin\phi_{k}}{\sigma} \right),  
 \end{split}
\label{eqn:gauss_hermite_functions}
\end{equation}

\noindent where $\sigma = 32^{\prime}/\sqrt{8 \ln 2}$ is the Gaussian width of the beam and $H_{i}$ and $H_{j}$ are Hermite polynomials. 
The dominant effects on the data quality will come from the lower order, $i+j\leq2$, terms for two reasons: (1) the higher-order terms are negligibly small and (2) we care most about leakage from local dipoles and quadrupoles, which will not average down as easily as the higher-order terms. A similar expansion can be done for the other beams, $\varepsilon_{Q}$, $\varepsilon_{U}$, $\xi_{QU}$, and $\xi_{UQ}$.

The lowest-order beam distortions can be written out explicitly, as 
\begin{equation}\begin{array}{rcccl}
f_{m0} \left( \theta, \phi \right) & \equiv & f_{00} & = & 1/(\sqrt{\pi \sigma^{2}}) \exp{\left( - \frac{\theta^{2}}{2 \sigma^{2}} \right)} \;, \\
f_{d1} \left( \theta, \phi  \right) & \equiv & f_{10} & = & \sqrt{2} (\theta \cos \phi / \sigma) f_{00} \;, \\
f_{d2} \left( \theta, \phi  \right) & \equiv & f_{01} & = & \sqrt{2} (\theta \sin \phi / \sigma) f_{00} \;, \\
f_{q1} \left( \theta, \phi  \right) & \equiv & f_{11} & = & (\theta^{2} \sin 2 \phi / \sigma^{2} ) f_{00} \;. \\
\end{array}
\label{eqn:gh_low_order1}
\end{equation}

\noindent We identify $f_{m0}$ as the monopole (scalar) beam,  $f_{d1}$ and  $f_{d2}$ as the horizontal and vertical dipoles, respectively, and $f_{q1}$ as the cross-shaped quadrupole. The plus-shaped quadrupole is defined to be
\begin{equation}
f_{q2} \equiv ( f_{20} - f_{02} ) / \sqrt{2} =  \frac{\theta^{2} \cos 2 \phi}{ \sigma^{2}} f_{00},
\label{eqn:gh_low_order2}
\end{equation}

\noindent and the differential width function is 
\begin{equation}
f_{m1} \equiv (f_{20} + f_{02} ) / \sqrt{2} = \left(\frac{\theta^{2}}{ \sigma^{2}} - 1 \right) f_{00}.
\label{eqn:gh_low_order3}
\end{equation}

\noindent We define $a$ coefficients that are normalized versions of the
$s$ coefficients in Eq.~\ref{eqn:lambda_expansion}:

\begin{equation}\begin{array}{rcl}
a_{m0} & \equiv & s^{\text{leak}}_{00} / s^{\text{main}}_{00} \;, \\
a_{m1} & \equiv & (1/\sqrt{2}) (s^{\text{leak}}_{20}+ s^{\text{leak}}_{02}) / s^{\text{main}}_{00} \;, \\
a_{d1} & \equiv & s^{\text{leak}}_{10} / s^{\text{main}}_{00} \;, \\
a_{d2} & \equiv & s^{\text{leak}}_{01} / s^{\text{main}}_{00} \;, \\
a_{q1} & \equiv & s^{\text{leak}}_{11} / s^{\text{main}}_{00} \;, \\
a_{q2} & \equiv & (1/\sqrt{2}) (s^{\text{leak}}_{20} - s^{\text{leak}}_{02}) / s^{\text{main}}_{00} \;. \\
\end{array}
\label{eqn:a_coeffs}
\end{equation}

Figure \ref{fig:model_IP_mono} summarizes the modeled ABS temperature-to-polarization leakage as a function of position on the focal plane; the color scale indicates the magnitude $\Lambda_{P}$. Figure \ref{fig:model_residual} shows the residuals as a function of position on the focal plane after removing the scalar, dipole, and quadrupole terms from the modeled leakage. Figure \ref{fig:model_outputs} shows the modeled monopole, dipole, and quadrupole leakage beams for $I \rightarrow Q$ and $I \rightarrow U$, derived respectively from $\lambda_{Q}$ and $\lambda_{U}$. Specifically, Figure \ref{fig:IQ_mono_leak} (\ref{fig:IU_mono_leak}) show $a_{m0}$ for $\lambda_{Q}$ ($\lambda_{U}$) for each position $k$ in the focal plane, which is equal to $\Lambda_{Q}$ ($\Lambda_{U}$). It should be noted that $a_{m0}$ for $\lambda_{Q}$ ($\lambda_{U}$) is nearly exactly $g_{I}^{4C}$ ($g_{I}^{4S}$) in Equation \ref{eqn:mod_mueller_2}, and this correspondence becomes exact as the beam width goes to zero. For later comparison to data, we construct an estimator of $\Lambda_{P}$ that is always positive in the model
\begin{equation}
\widetilde{\Lambda}_{P} = - \Lambda_{Q} \cos 2\phi - \Lambda_{U} \sin 2\phi
\label{eqn:signed_leakage_combo}
\end{equation}

\noindent and an orthogonal function that is always zero
\begin{equation}
\widetilde{\Lambda}^{\bot} = \Lambda_{Q} \sin 2\phi - \Lambda_{U} \cos 2\phi .
\label{eqn:ortho_signed_leakage_combo}
\end{equation}
Here $\phi = \phi_{k0}$ when $\widetilde{\Lambda}_{P}$ or
$\widetilde{\Lambda}^{\bot}$ is calculated for a detector $k$.

Figures \ref{fig:IQ_dip_leak} and \ref{fig:IU_dip_leak} show the magnitude of the two dipole terms $\sqrt{a_{d1}^{2} + a_{d2}^{2}}$ in the color scale with the direction of the dipole indicated by the overplotted arrows, for $\lambda_{Q}$ and $\lambda_{U}$, respectively. Finally, Figures \ref{fig:IQ_quad_leak} and \ref{fig:IU_quad_leak} show the magnitude of the two quadrupole terms $\sqrt{a_{q1}^{2} + a_{q2}^{2}}$ in the color scale with the direction of the quadrupole indicated by the overplotted crosses for $\lambda_{Q}$ and $\lambda_{U}$, respectively. 
We do not plot the coefficient for the differential width function defined in Equation \ref{eqn:gh_low_order3} as it is zero everywhere.

\section{Measurement of Scalar Leakage}
\label{sec:scalar}
We use $A(\chi)$, defined in Equation \ref{equ:a_of_chi}, to
characterize the leakage $\Lambda_{P}$ (Equation
\ref{eqn:total_leak_def}). The function $A(\chi)$ is measured for each $\sim$one hour
constant-elevation scan (CES) of the CMB.
The terms with $\cos 4\chi$ and $\sin 4\chi$ dependence can be written as
\begin{equation}
 A(\chi) = \left( \Lambda_{Q} I + A_0^{Q} \right) \cos 4\chi
 + \left( \Lambda_{U} I + A_0^{U} \right) \sin 4\chi +
  \cdots \:. 
\end{equation}

Note that we have switched to beam-averaged quantities in this
equation. To determine $\Lambda_{Q}$ and $\Lambda_{U}$, we use the fact
that the sky intensity $I$ changes with precipitable water vapor (PWV). We estimate the atmosphere contribution to the intensity for each CES from the elevation angle of the ABS telescope and the PWV measured by APEX.~\footnote{http://www.apex-telescope.org/weather/} To convert to temperature units, we use the ABS frequency bandpasses presented in our previous publication\cite{2014SPIE.9153E..0YS} and the ATM (atmospheric transmission at microwaves)
model\cite{2001ITAP...49.1683P} implemented as the AATM package.\footnote{https://www.mrao.cam.ac.uk/~bn204/alma/atmomodel.html}

\begin{figure}[tbp]
	\begin{center}
		\includegraphics[width=0.45\textwidth]{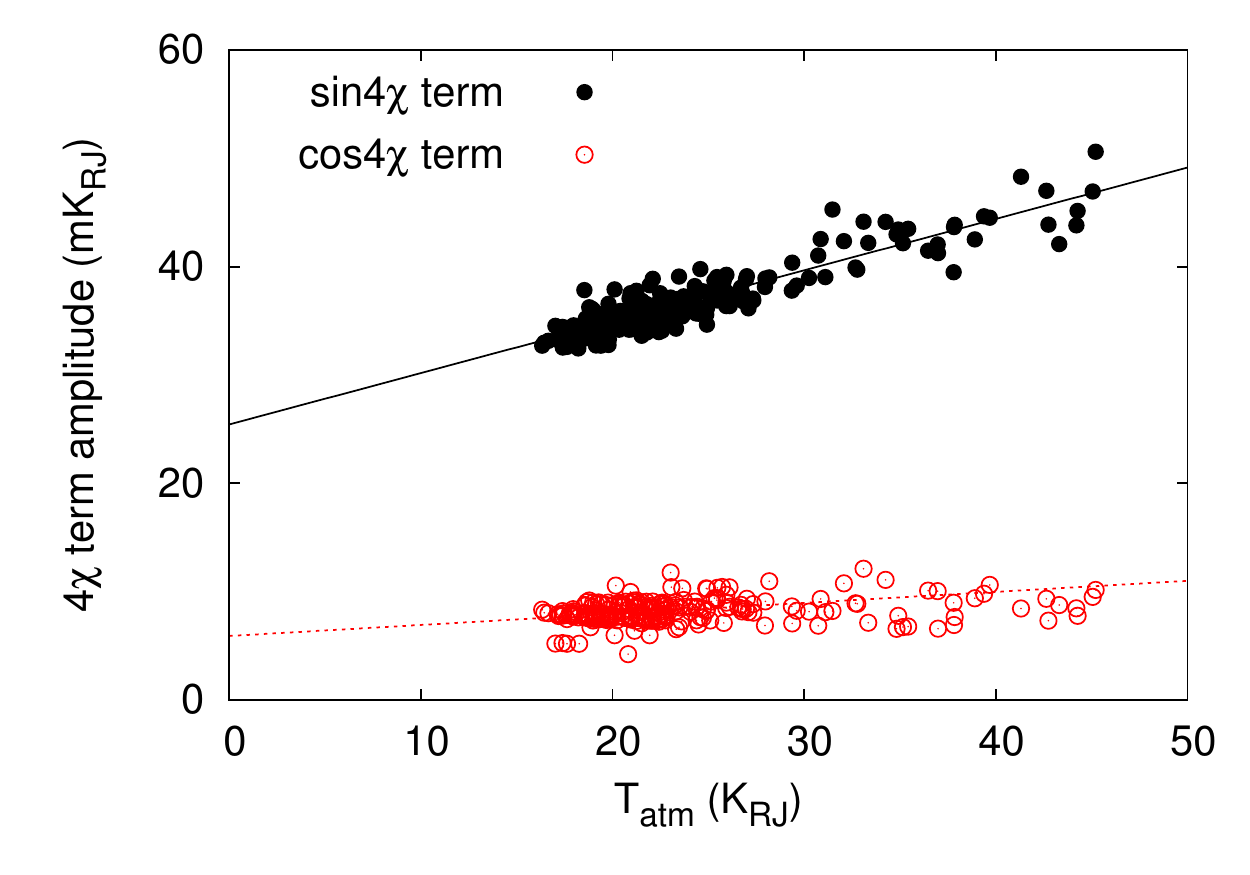}
		\caption{\label{fig:4f_vs_tsky} %
			An example of the relation between the brightness temperature of the atmosphere and the amplitudes of the $\cos 4\chi$ and $\sin 4\chi$ terms in $A(\chi)$ for one of the ABS detectors. Each point in this plot corresponds to an $\sim$ hour-long CES. Note that we estimate the coefficients in temperature units (without converting to the CMB units appropriate for differential measurements near the peak of that blackbody).
		}
	\end{center}
\end{figure}

Figure \ref{fig:4f_vs_tsky} shows an example of such a relation between the sky intensity $I$, and the amplitudes of the $\cos 4\chi$ and $\sin 4\chi$ terms. The slopes in this correlation plot correspond to the leakage coefficients $\Lambda_{Q}$ and $\Lambda_{U}$. We evaluate the statistical uncertainty by dividing the data into subsets, each corresponding to a period of a few weeks to months. The estimated leakage coefficients are consistent among the data subsets. From the variance of these subset estimates, we estimate the uncertainty on $\Lambda_{P}$ for each detector as 0.006\% (0.007\%) for group A (B) detectors. 

\begin{figure}[t]
 \begin{center}
  \includegraphics[width=0.45\textwidth]{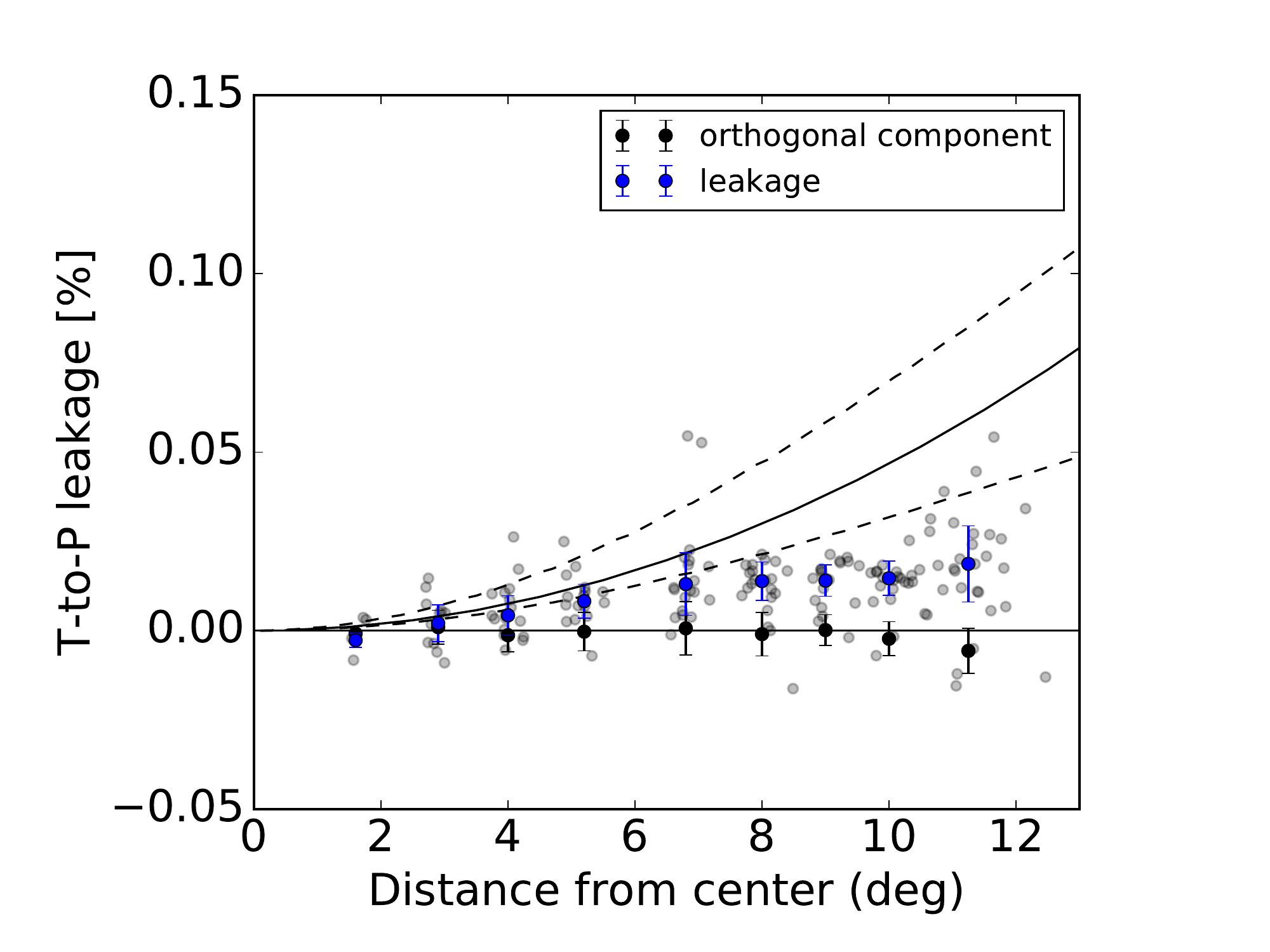}
  \caption{  Intensity-to-polarization leakage, estimated using the $A (\chi)$ signal versus precipitable water vapor (PWV) described in the text, versus radial distance from the boresight for group A detectors. These detectors were chosen for this analysis, because they have well-measured and similar bandpasses. Each gray dot is the leakage $\widetilde{\Lambda}_{P}$ (defined in Equation \ref{eqn:signed_leakage_combo}) averaged over two seasons of observations for a single detector. We estimate the mean error for each of these points to be 0.006\%. The blue dots are binned averages for groups of detectors at approximately the same focal plane radius, with the error bar the variance of the group. The black dots show the binned averages for $\widetilde{\Lambda}^{\bot}$, which are close to zero as predicted by the model. The solid black curve is the fiducial estimate from the transfer-matrix model with the dashed lines corresponding to uncertainty in the anti-reflection coating thickness of $\pm 25$~$\mu$m due to manufacturing tolerances.} 
\label{fig:leak_v_rad}
  \end{center}
\end{figure}

Next, we compute $\tilde{\Lambda}_P$ (Eq.~\ref{eqn:signed_leakage_combo}) from
the measured $\Lambda_{Q}$ and $\Lambda_{U}$
and compare it to the estimates from the transfer-matrix model.
As shown in Figure \ref{fig:leak_v_rad},
the model correctly predicts that
$\widetilde{\Lambda}_{P}$ increases with distance from boresight
and that the orthogonal leakage
 $\widetilde{\Lambda}^{\bot}$ is consistent with zero.
 The data values of $\widetilde{\Lambda}_{P}$ are not forced
 to be positive and can in principle take negative values.
 The model also predicts the amplitude of
 $\widetilde{\Lambda}_{P}$ within a factor two.
This overall agreement between the model prediction and the ABS data
demonstrates that the transfer-matrix model will be a useful tool in
designing future experiments.
 %
The factor two difference in amplitude could be due to uncertainties in the thickness, index of refraction, or absorptive loss of the HWP and its anti-reflection (AR) coating. For instance, the blue band in the figure shows the effect on the modeled leakage for the expected maximum $\pm$25 $\mu$m uncertainty in the anti-reflection coating thickness due to manufacturing tolerances. Similar levels of uncertainty can be attributed to thickness variation of the adhesive layer between the HWP and its anti-reflection coating.

\begin{figure}[htbp]
	\begin{center}
		\includegraphics[width=0.45\textwidth, clip=true, trim=0cm 0cm 0cm 1.5cm]{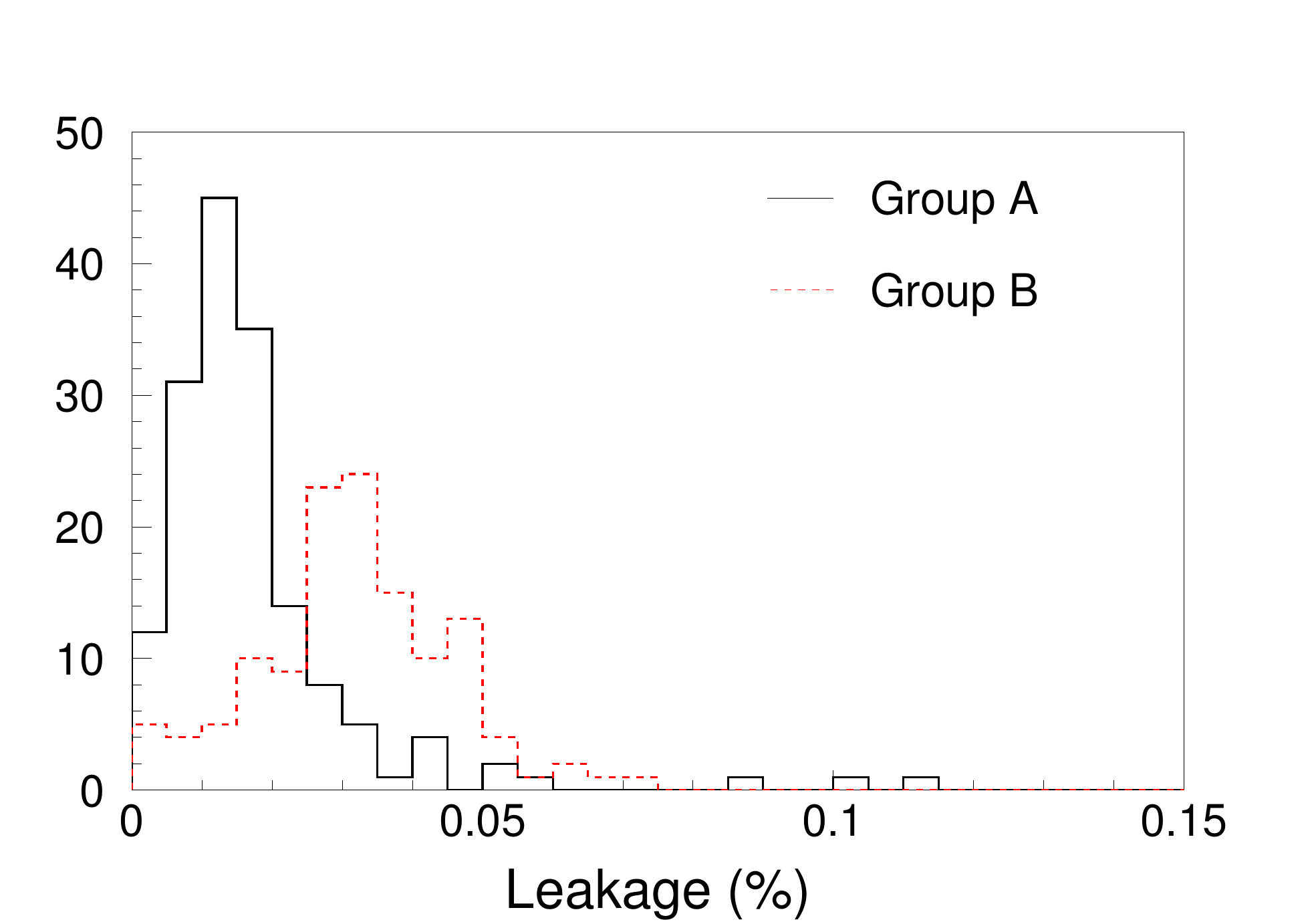}
		\caption{\label{fig:leakage_histo} %
			A histogram of the $T \rightarrow P$ leakage, defined as
	 $\sqrt{{\Lambda_Q}^2 + {\Lambda_U}^2}$.
	 Each entry of this histogram corresponds to a detector.
	 We note that this is a biased estimator due to the statistical error of the measurement; the level of the bias is estimated to be 0.009\% (0.011\%) for the group A (B) detectors. The median value of the leakage is 0.013\% (0.031\%) for group A (B).
		}
	\end{center}
\end{figure}

Figure~\ref{fig:leakage_histo} presents a histogram of the leakage
coefficients $\Lambda_P \equiv \sqrt{{\Lambda_Q}^2 + {\Lambda_U}^2}$.
Note that if every $\Lambda_{Q}$ and $\Lambda_{U}$ were drawn from distributions with zero mean and width $\sigma$, the histogram would not peak at zero. After correcting for this bias, the median value of the total leakage is $\Lambda_{P} = 0.013\%$ (0.031\%) for the detector group A (B). We put a conservative upper limit on the effective scalar leakage in ABS of
\begin{equation}
\Lambda_{P} < 0.03\% \hspace{10pt}(\text{magnitude method})
\end{equation}
from these data, noting that the group A detectors dominate the statistical weight. 
The typical value of the scalar leakage, 0.013\%, is considerably smaller than has been previously reported, as indicated in Table \ref{tbl:other_experiments}.


Although estimating an upper limit on the leakage from estimates of $\Lambda_{P}$ is the more conservative approach, we can also provide a more direct estimate that circumvents the problem of bias in estimating a magnitude. We investigate the mean values of the leakage coefficients across the focal plane, finding $\bar{\Lambda}_{Q}$ = 0.005\% ($-0.001$\%) and $\bar{\Lambda}_{U}$ = 0.007\% ($-0.003$\%) for detectors from group A (B). We estimate the errors on these means as 0.003\% for each case from the standard deviations of the distributions of $\Lambda_{Q}$ and $\Lambda_{U}$ for each detector group. We use the $2\sigma$ limit on $\sqrt{\left(\bar{\Lambda}_{Q}\right)^{2} + \left(\bar{\Lambda}_{U}\right)^{2}}$ from group A to provide a second estimate of the effective scalar leakage:
\begin{equation}
\Lambda_{P} < 0.014\% \hspace{10pt}(\text{mean value method})
\end{equation}

Neglecting possible
suppression of the systematic error by sky rotation
and cancellation in averaging pixels across the focal
plane,
a leakage coefficient of 0.014\% leads to systematic bias
as small as
a tensor-to-scalar ratio $r$ of 0.002--0.003
by taking the formalism of Shimon
et. al. 2008\cite{2008PhRvD..77h3003S}.
%
%
In order to assess the level of systematic-error mitigation
from sky rotation and focal plane averaging, we perform an end-to-end
pipeline simulation of ABS for season 1 and 2 observations,
based on the detector-by-detector leakage coefficients predicted by our
model (Fig.~\ref{fig:model_outputs}).
Figure~\ref{fig:syst_power_spec} shows the resultant estimate of the
systematic bias.  The bias is at a level of $r < 0.001$ for angular scales of $\ell < 100$.
Sky rotation and the cancellation
across the focal plane yield factors of $\sim 2$
and $\sim 5$, respectively, reduction of the systematic error in power.
  There is significant room for improvement in future experiments, because
  (1) ABS had a scan pattern that was not optimal for reducing systematic error via sky rotation,
  specifically for the monopole leakage;
  (2) ABS had a focal plane with highly non-uniform sensitivity, degrading
  the focal-plane cancellation; and (3) the ABS scan width was $\sim 7$\,degrees,
  smaller than the focal plane diameter of $\sim 20$\,degrees, again
  decreasing cancellation from focal-plane averaging.
We also note that the average modeled scalar leakage is
$\sim 0.022$\%, while our data implies a smaller amplitude of 0.014\%
as can be seen in Fig.~\ref{fig:leak_v_rad}.
If we scale the leakage by this factor,
the bias on $r$ would decrease by a factor of $\simeq 2.5$
compared to Fig.~\ref{fig:syst_power_spec}.

\section{Higher-Order Leakage: Dipole and Quadrupole}
\label{sec:higher}
We now constrain the higher order terms in $\lambda_{Q}(\theta,\phi)$ and $\lambda_{U}(\theta, \phi)$.  
All the terms are consistent with zero, and thus we derive upper limits
on the leakage coefficients $a_d$, $a_q$, and $a_{m1}$ based on the
errors in our measurement.
These constraints are obtained by making maps of Jupiter.
An ideal polarization modulator without any polarization systematics
would lead to a null signal in these polarization maps assuming Jupiter
is unpolarized.
For a non-ideal HWP, the angle-dependent leakage beams $\lambda_{Q}(\theta, \phi)$ and $\lambda_{U}(\theta, \phi)$ will appear in maps of an unpolarized source.  Scalar leakage shows up as a spurious point source at the location of the unpolarized source, with the same shape as the total intensity beam shape. Higher-order terms show up as zero-mean patterns (e.g., dipole or quadrupole) in the maps. 

We note that our constraints are robust against possible intrinsic
polarization of Jupiter for two reasons.  First, to a good approximation, polarization of
Jupiter would only contribute to the monopole terms
and the higher order terms are immune to the polarization of Jupiter.
Second, we only put upper limits on
the leakage coefficients and thus contribution from non-zero polarization of
Jupiter would only make the limits more conservative.

\begin{figure*}[htbp]
	\includegraphics[width=0.325\textwidth]{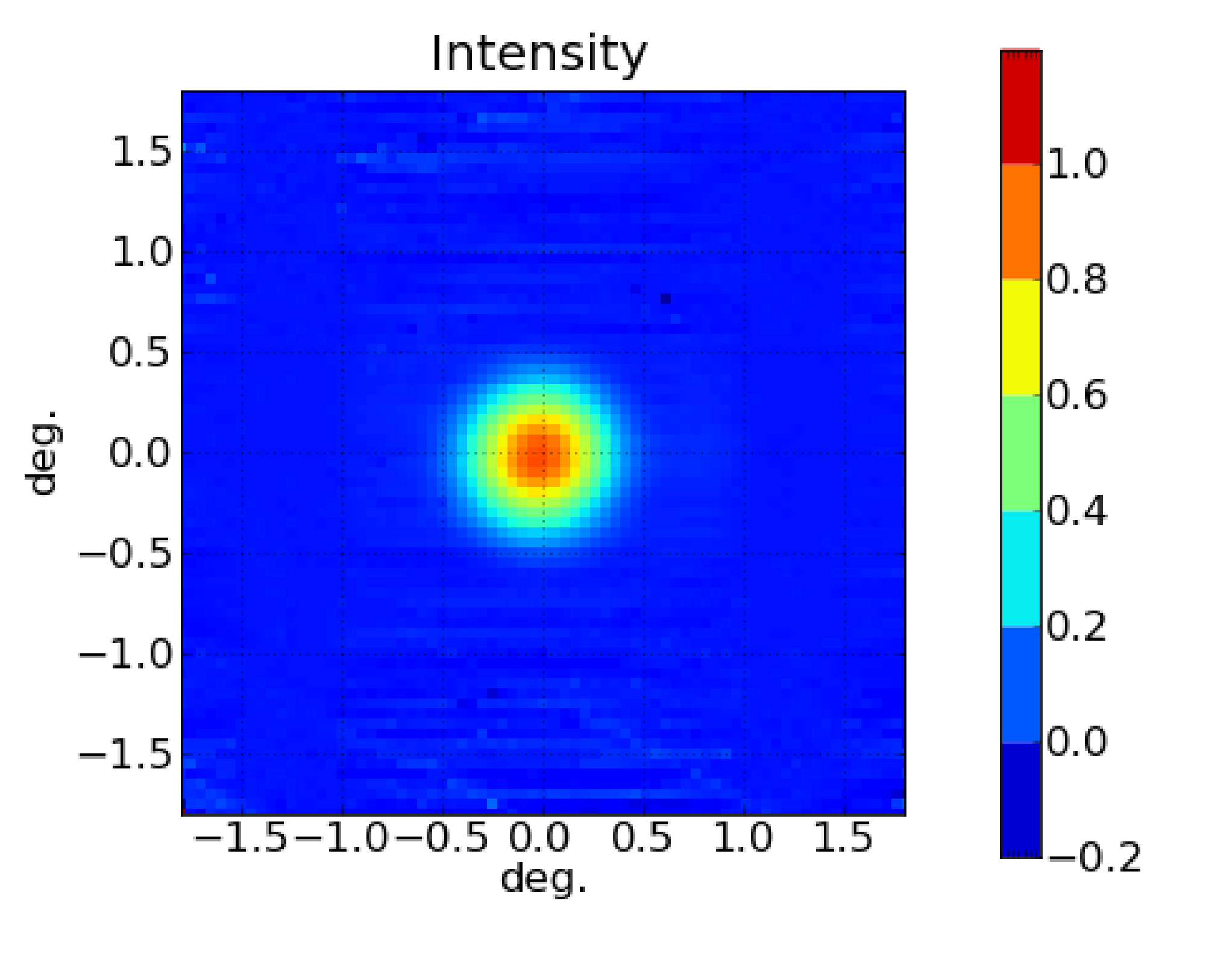}
	\includegraphics[width=0.325\textwidth]{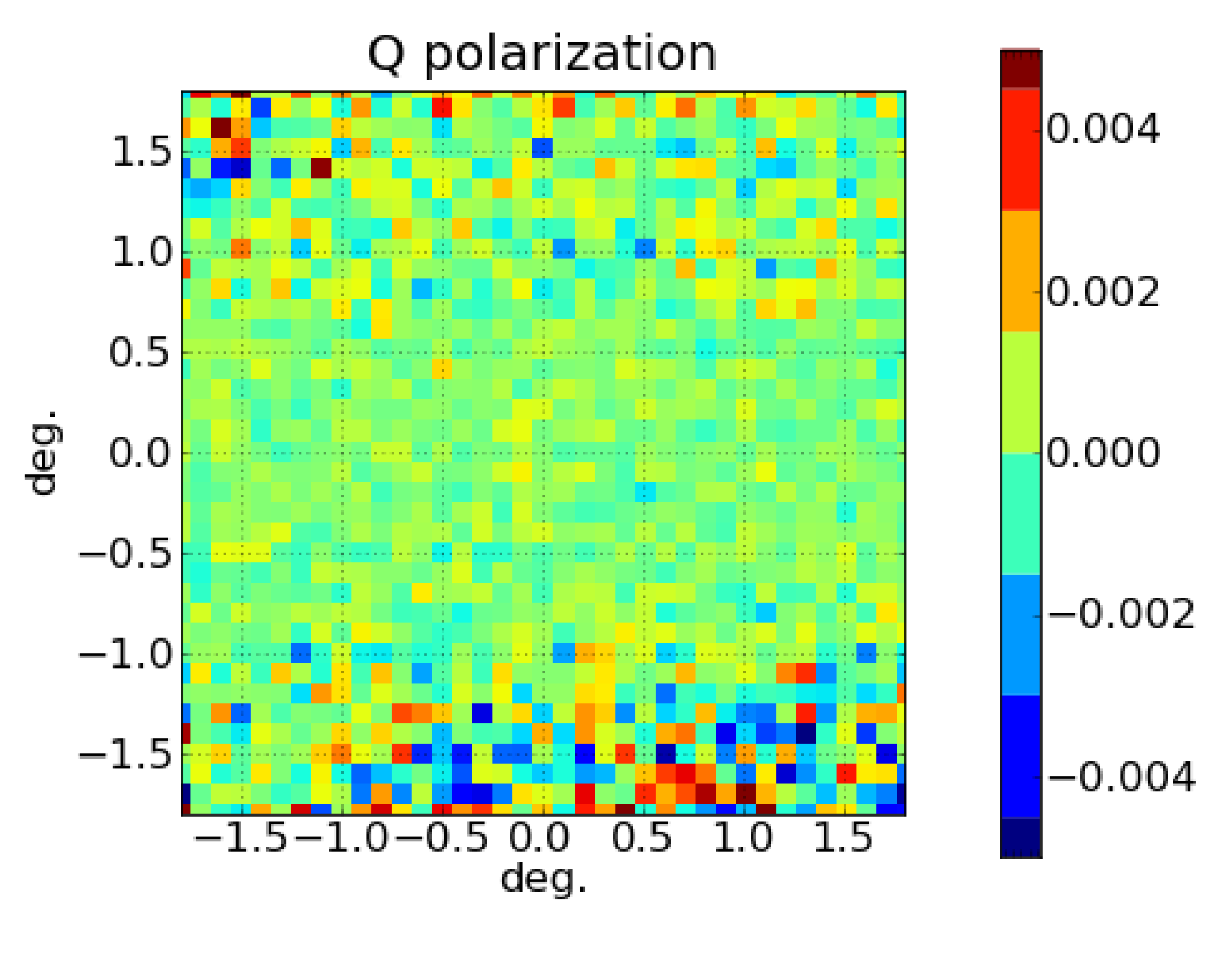}
	\includegraphics[width=0.325\textwidth]{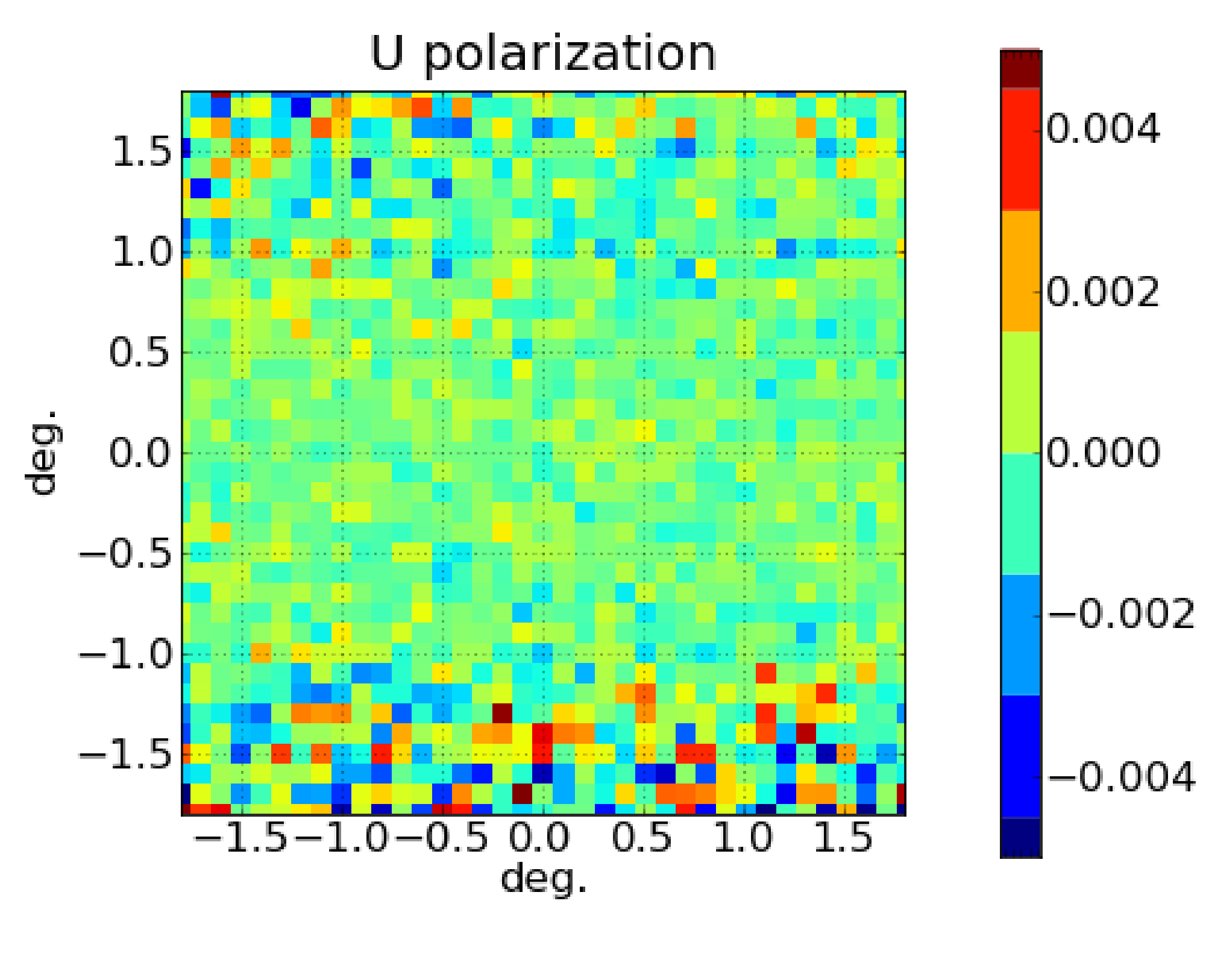}
	\caption{\label{fig:maps} %
		Stacked intensity and ``polarization'' maps of Jupiter for pod 4 detectors, one of the best-observed pixel groups.
		Left: intensity map of Jupiter, corresponding to the
 ABS intensity beam.  Center and Right: Q and U ``polarization'' maps of Jupiter using demodulated data. 
 The color scale is normalized such that the peak of the intensity map
 (left) is unity.
     The maps are for a group of 20 detectors near the center of the focal plane.
 Assuming Jupiter is unpolarized, we expect no signal when the HWP creates no spurious polarization.
 Note that these maps do not constitute a limit on the possible
 polarization of Jupiter; $\sim 30$ observations are coadded to create
 the maps in a manner to coherently add up possible
 instrumental polarization, but not necessarily to add the polarization
 of Jupiter.
	}
\end{figure*}

For detectors from group A, we make stacked maps of multiple Jupiter observations for sets of ten neighboring pixels (20 detectors).  
Figure~\ref{fig:maps} presents an example of Jupiter maps for one of the best-calibrated sets (hereafter ``pod 4''). This group is in a region $2^{\circ}-6^\circ$ from the center of the focal plane. 
To create these maps, some $\sim 30$ observations are stacked in
co-azimuth vs. co-elevation coordinates, which coherently add up
the $T \rightarrow P$ leakage beam but not necessarily add up possible intrinsic polarization
of Jupiter.
The left panel shows the total intensity beam. The center and right panels show the demodulated data maps. The maps are normalized such that the peak of the intensity map is unity.  No spurious polarization is evident. We calculate two types of radial projections of the polarization maps by integrating over $\cos n\phi$ or $\sin n\phi$ to pick out  dipole ($n=1$) and quadrupole ($n=2$) terms:
\begin{eqnarray}
Q_n^{c[s]}(\theta) &\equiv& \int \mathrm{d}\phi Q(\theta,\phi) \cos n\phi [\sin
n\phi] \label{equ:radial_average_Q} \\ 
U_n^{c[s]}(\theta) &\equiv& \int \mathrm{d}\phi U(\theta,\phi) \cos n\phi [\sin
n\phi] \label{equ:radial_average_U}
\end{eqnarray}
We note that the quadrupole terms are irreducible by sky rotation~\cite{2003PhRvD..67d3004H} and thus among the most important sources of systematic error.  Among the four quadrupole terms, $Q^s_2(\theta)$ and $U^c_2(\theta)$ lead to spurious $B$ modes, while $Q^c_2(\theta)$ and $U^s_2(\theta)$ result in spurious $E$ modes.

To compare these data with the prediction presented in
Section~\ref{sec:beam_decomp}, we define the following radially-projected leakage beam templates for two monopoles ($f_{m0}$ and $f_{m1}$), dipole ($f_{d}$) and a quadrupole ($f_q$):
\begin{equation}
  \label{equ:def_monopole_quadrupole}
 \begin{split}
 f_{m0}(\theta) & \equiv \frac{\sqrt{\pi\sigma^2}}{2\pi} \int \mathrm{d}\phi
  f_{m0}(\theta,\phi) \:,
  \\
 f_{m1}(\theta) & \equiv \frac{\sqrt{\pi\sigma^2}}{2\pi} \int \mathrm{d}\phi
  f_{m1}(\theta,\phi) 
  \:,
  \\
 f_{d}(\theta) & \equiv \frac{\sqrt{\pi\sigma^2}}{2\pi} \int \mathrm{d}\phi
  \cos \phi \: f_{d1}(\theta,\phi)
  \:,
  \\
 f_{q}(\theta) & \equiv \frac{\sqrt{\pi\sigma^2}}{2\pi} \int \mathrm{d}\phi
  \cos 2\phi \: f_{q2}(\theta,\phi)
  \:.
 \end{split}
\end{equation}
The prefactors $\sqrt{\pi\sigma^2}$ are normalization factors such that
the peak of the main beam is unity in the maps shown in
Fig.~\ref{fig:maps}. That is, this normalization allows us to directly
obtain coefficients $a_{m0}$, $a_{m1}$, $a_{d1}$, $a_{d2}$, $a_{q1}$ and
$a_{q2}$ by fitting the radially-averaged leakage maps
(Eqs.~\ref{equ:radial_average_Q} and \ref{equ:radial_average_U}) to these templates. For $P = Q, U$, fitting $f_d$ on $P_1^c$ ($P_1^s$) yields $a_{d1}$ ($a_{d2}$), and fitting $f_q$ on $P_2^c$ ($P_2^s$) yields $a_{q1}$ ($a_{q2}$). In the following, we treat $a_{d1}$ and $a_{d2}$ together as $a_d$ and $a_{q1}$ and $a_{q2}$ together as $a_q$, unless otherwise noted. 
We fit the templates and determine the coefficients one
by one.

\begin{figure*}[htbp]
 \includegraphics[width=0.32\textwidth]{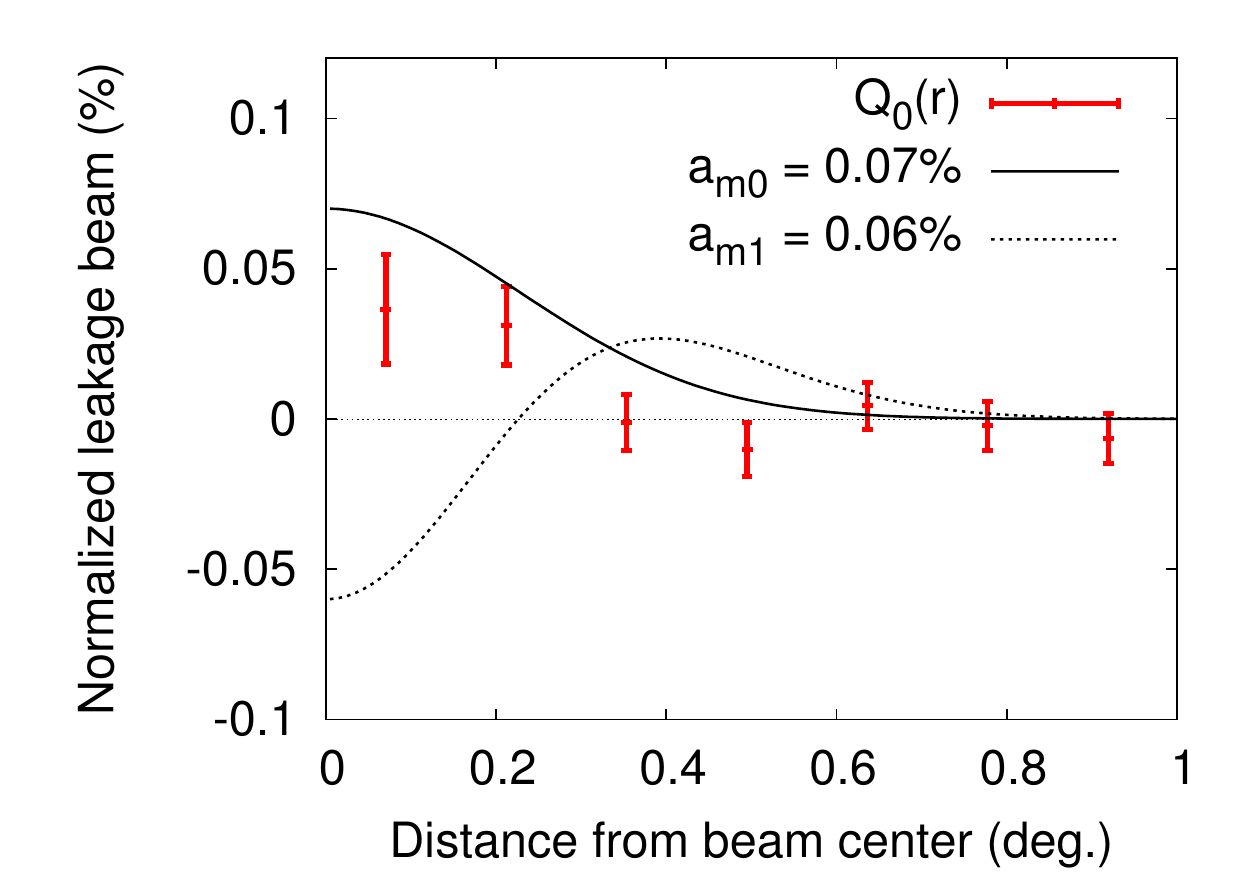}
 \includegraphics[width=0.32\textwidth]{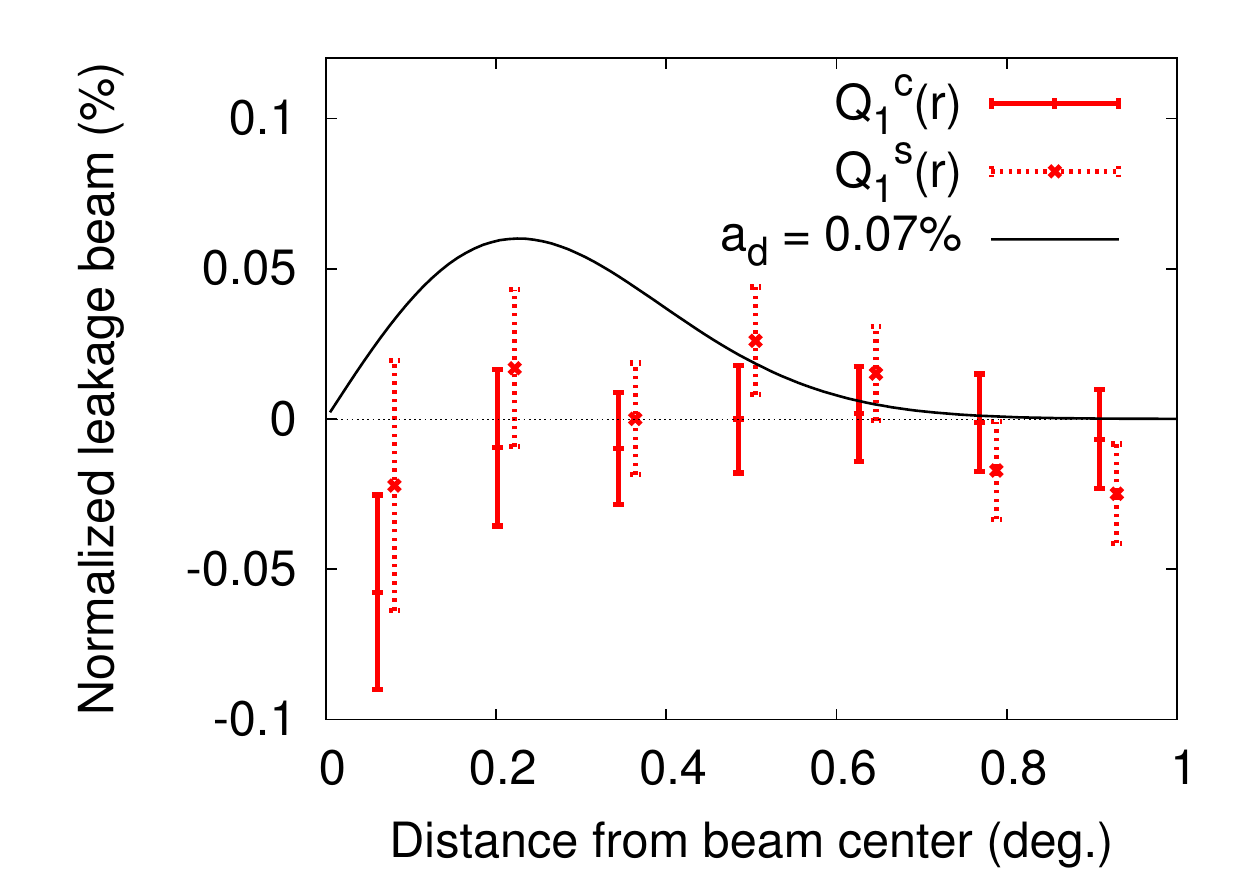}
 \includegraphics[width=0.32\textwidth]{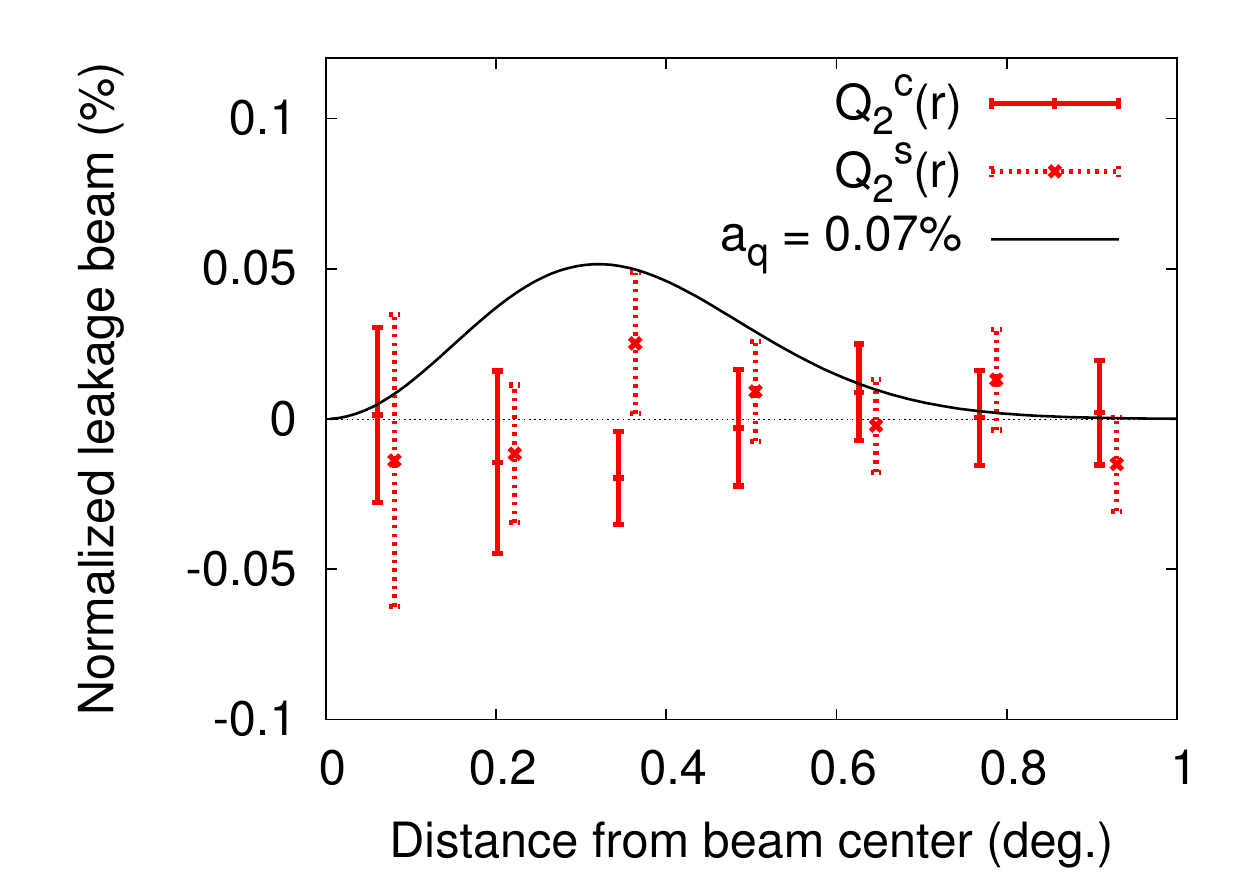}
 \caption{\label{fig:radial_averages} %
 The radial averages of monopole, dipole, and quadrupole
 in Q polarization 
 defined by Eq.~(\ref{equ:radial_average_Q})
 for the pod 4 maps shown in Fig.~\ref{fig:maps}, center.
 Plotted curves correspond to example templates for
 monopole, dipole, and quadrupole terms
 with amplitudes that correspond to the upper limits shown in Table~\ref{tab:syst_error_limits2}.
 %
 }
\end{figure*}

Figure~\ref{fig:radial_averages} shows the monopole,
dipole and quadrupole leakage components derived from the maps shown in
the center and right panels of Figure~\ref{fig:maps} following
Equations~\ref{equ:radial_average_Q} and \ref{equ:radial_average_U}.
The data are
consistent with zero at the $\sim\, 0.02$\% level. The curves for templates are
shown for comparison.

We note that these leakage beam templates are equivalent to those defined in Shimon, et al. 2008,~\cite{2008PhRvD..77h3003S} in describing a ``two-beam'' experiment which required beam differencing, in the limit of small leakages. Three of the four coefficients for the templates correspond to differential gain ($g$), differential pointing ($\rho$) and differential ellipticity ($e$) as $a_{m0} = g/2$, $a_d = \rho/(\sqrt{8} \sigma)$  and $a_q = e $. The differential beam width ($\mu$) and the coefficient $a_{m1}$ do not
have the same template function shape. Further details are in Appendix \ref{sec:gh_to_diff}.

While none of the 12 amplitudes\footnote{%
We fit six amplitudes (Eq.~\ref{eqn:a_coeffs}) for each of the Q and U leakage maps,
yielding 12 amplitudes in total.
} fitted show significant deviation from
zero,  the total $\chi^2$ to zero for the 12 amplitudes is 23 with 12
degrees of freedom.
We consider this
excess is likely to be because of an underestimate of the errors, which
are empirically obtained.  We thus conservatively place upper limits on
these amplitudes using errors inflated by a factor of
$\sqrt{23/12}=1.4$.
The ``pod 4'' column in Table~\ref{tab:syst_error_limits2} shows the 2-$\sigma$
upper limits.
We repeat the same analyses on another eight (of twelve) relatively well-calibrated groups of batch A detectors; the eight groups broadly distribute across the top half of the focal plane.
Using the scatter among the eight groups, we estimate
the error per group.  We find no significant deviation from zero, with a
total $\chi^2$ of 10.9 for 12 degrees of freedom.
Thus, we put upper limits on these coefficients without inflating the
errors here.
The ``others'' column in 
 Table~\ref{tab:syst_error_limits2} summarizes those limits.
We note that these limits are more conservative than a usual
2-$\sigma$ limit in that we take the worst case among two (four)
2-$\sigma$ upper limits for monopole (dipole and quadrupole) amplitudes.

\begin{figure}[htbp]
 \begin{center}
  \includegraphics[width=0.45\textwidth]{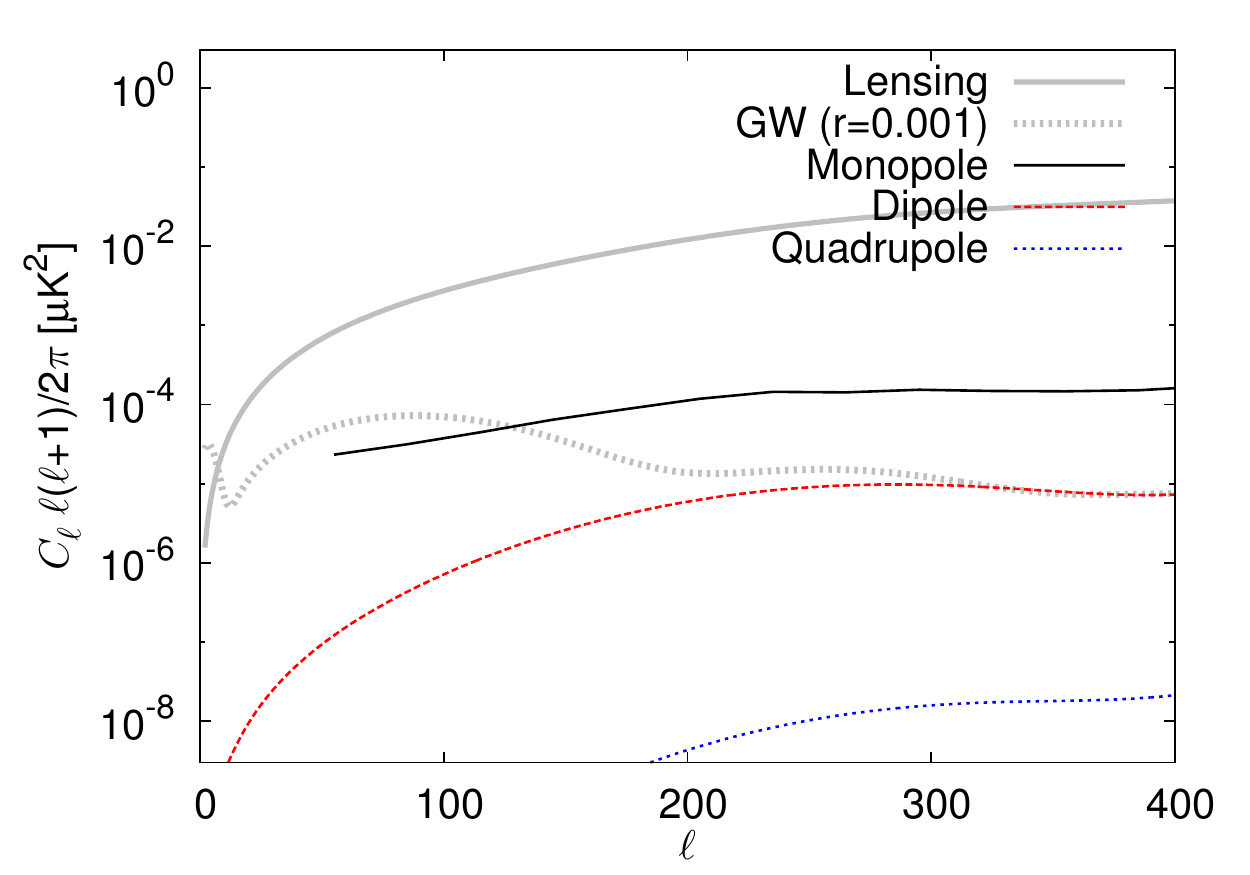}
  \caption{\label{fig:syst_power_spec} %
  The level of $I \rightarrow P$ leakage systematic error for monopole, dipole,
  and quadrupole terms compared with the $BB$ power spectra of lensing
  and primordial gravitational waves (GW) with a tensor-to-scalar ratio
  $r=0.001$.  The leakage coefficients are taken from the model shown in
  Fig.~\ref{fig:model_outputs}.
  We emphasize that these are the estimates before any correction (e.g.,
  deprojection).
  For the monopole, we perform an end-to-end pipeline simulation of ABS for
  season 1 and 2 observations, yielding a level of systematics below
  $r=0.001$ for $\ell<100$.
  Note that the average amplitude of the coefficients from our model is
  $\sim 0.022$\%, while our data implies a smaller amplitude of 0.014\%
  as can be seen in Fig.~\ref{fig:leak_v_rad}.
  Taking the smaller amplitude implied by our data leads to a reduction in systematic error by a factor $\simeq 2.5$.
  For the dipole and quadrupole leakages, we take typical leakage
  values of $2.9 \times 10^{-5}$ and $1.4 \times 10^{-6}$, respectively,
  from the model (Fig.~\ref{fig:model_outputs})
  and analytically calculate the systematic
  bias;\cite{2008PhRvD..77h3003S} we neglect possible mitigation of the dipole leakage by sky rotation
  and focal plane cancellation. The bias is well below the levels of $r=0.001$ or the
  gravitational lensing B modes.
  }
 \end{center}
\end{figure}

Taking the formalism of Shimon
et. al. 2008\cite{2008PhRvD..77h3003S} and
neglecting possible systematic-error mitigation by sky rotation,
we relate these upper limits to the systematic bias in B-mode power.  
The upper limits of the quadrupole and differential width terms
correspond to $r = 0.001$ or lower.  The limit on the dipole term
corresponds to $r \sim 0.003$ (0.01) for pod 4 (others).  With an optimal
scan strategy, sky rotation would mitigate the dipole systematics 
further.  We also note that these limits are dominated by the
uncertainties of our beam measurement, and the true values of the
coefficients are expected to be lower.
Figure~\ref{fig:syst_power_spec} shows the expected bias 
based on typical
values of the leakage coefficients from our model (Fig.~\ref{fig:model_outputs}),
which are $2.9 \times 10^{-5}$ and $1.4 \times 10^{-6}$
for the dipole and quadrupole leakages, respectively.
Here we neglect possible mitigation of the dipole leakage by sky rotation
  and focal plane cancellation.
The bias is well below the level of a primordial gravitational wave signal with $r=0.001$ or gravitational lensing B modes.

%
%
   %
   %
   
  \begin{table}[htbp]
  \caption{ \label{tab:syst_error_limits2} %
   Upper limits on the amplitudes of measured $I \rightarrow P$ leakage terms from analysis of stacked maps of Jupiter.   
   Each row corresponds to a group of two (four) amplitudes for monopole (dipole and quadrupole) terms; the worst case of the two (four) 2-$\sigma$ upper limits are shown here.
   For the second monopole term, we present constraints on both $a_{m1}$ (see Equation~\ref{equ:def_monopole_quadrupole}) and the two-beam\cite{2008PhRvD..77h3003S} parameter $\mu$ since their template functions are different.
   The Pod 4 stacked maps are more sensitive than the ``Others'' maps,
   as explained in the text.  Note that Section~\ref{sec:scalar}
   provides a measurement of the scalar leakage which is more
   constraining than the limits in this table.
   }
    \begin{tabular}{p{7em}p{9em}p{4em}}
     \hline \hline
    & Pod 4 & Others \\
     \hline 
     \rule[-2mm]{0mm}{2mm}
     $|a_{m0}|$ (\%) & $<0.07$ & $< 0.13$ \\
     \rule[0mm]{0mm}{2mm}
     $|a_{m1}|$ (\%) & $< 0.06$ & $< 0.09$ \\
     \rule[-2mm]{0mm}{2mm}
     $\quad |\mu|$ (\%) & $< 0.05$ & $< 0.09$ \\
     \rule[-2mm]{0mm}{2mm}
     $|a_{d}|$ (\%) &  $< 0.07$ & $< 0.13$ \\
     \rule[-2mm]{0mm}{2mm}
     $|a_{q}|$ (\%) &  $< 0.07$ & $< 0.14$ \\
     \hline
    \end{tabular}
  \end{table}

  \begin{table}[bp]
  	\caption{ \label{tbl:other_experiments} %
  		Published estimates of $I \rightarrow P$ leakage for various experiments. When quoted for two-beam experiments, the leakage estimates are converted to the Gauss-Hermite basis used here according to the prescriptions of Appendix \ref{sec:gh_to_diff}. 
  For $|a_{m0}|$, we present the median value from measurements of $A(\chi)$ as detailed in Section \ref{sec:scalar}.
   Upper limits on higher-order beam terms are derived from Jupiter maps from Section \ref{sec:higher}.
   }
  	\begin{tabular}{lc@{\hskip 5pt}c@{\hskip 10pt}c@{\hskip 10pt}c} 
  		\hline \hline
  		& $|a_{m0}|$ & $|\mu|$ & $|a_{d}|$ & $|a_{q}|$ \\
  		& (\%) & (\%) & (\%) & (\%) \\
  		\hline  
  		\rule[0mm]{0mm}{2mm}
  		BICEP\cite{2010ApJ...711.1123C} & $<1.1$ & --- & $0.5\pm0.1$ & --- \\
  		\rule[-2mm]{0mm}{2mm}
  		BICEP2/Keck\cite{2015ApJ...806..206B} & 1 & $0.08\pm0.4$ & $0.64\pm1.7$ & $0.7\pm1.2$ \\
  		\rule[-2mm]{0mm}{2mm}
  		MAXIPOL\cite{2007ApJ...665...42J} & 1--4 & --- & --- & --- \\
  		\rule[-2mm]{0mm}{2mm}
  		QUIET Q\cite{2011ApJ...741..111Q} & 0.2--1 & --- & 0.1 & 0.1 \\ 
  		\rule[-2mm]{0mm}{2mm}
  		QUIET W\cite{2012arXiv1207.5562Q} & 0.4 & --- & 0.4 & 0.3 \\ 
  		\rule[-2mm]{0mm}{2mm}
  		WMAP\cite{2007ApJS..170..335P} & 0.1 & 6--8 & --- & --- \\ \hline
  		\rule[-2mm]{0mm}{2mm}
  		\textbf{ABS} &$0.013$  & $<0.05$ & $< 0.07$ & $< 0.07$ \\
  		\hline
  	\end{tabular}
  \end{table}

\section{Conclusion}
\label{sec:conclusion}
We have demonstrated low temperature-to-polarization systematic errors from a continuously-rotating HWP in observing CMB polarization with the ABS instrument.
Table~\ref{tbl:other_experiments} compares leakage results from ABS with
other CMB experiments.
Estimated levels of the ABS errors are presented using a transfer-matrix model.\cite{2013ApOpt..52..212E} The scalar leakage component is measured to be consistent with expectations, and we put a conservative upper limit on its magnitude of 0.01--0.03\%.
The model correctly predicts two trends found in the
data: the increase of the leakage as a function of the distance of a pixel 
from the center of the focal plane,
and the relation between the direction of the leakage polarization
and the position of a pixel in the focal plane.
The higher-order dipole and quadrupole terms are not detected, leading to upper limits on each of 0.07\%. This is also consistent with expectations. 
Before any systematic error mitigation due to cross-linking or boresight
rotation, the upper limits correspond to $r \lesssim 0.01$.
The measured scalar leakage and the theoretical level of 
 dipole and quadrupole leakage produce systematic error 
 of $r < 0.001$ for the ABS survey and focal-plane layout before 
 any data correction such as so-called deprojection.
Our study demonstrates the benefits of using a HWP for systematic error mitigation and the value of the transfer-matrix model as a tool for designing future experiments.



\section*{Acknowledgments}
Work at Princeton University is supported by the U.S. National Science
 Foundation through
awards PHY-0355328 and PHY-085587, the U.S. National Aeronautics and
 Space Administration (NASA) through award NNX08AE03G,
the Wilkinson Fund, and the Mishrahi Gift.  
%
Work at NIST is supported by the NIST Innovations in
Measurement Science program. 
Work at LBNL is supported by the U.S. Department of Energy, Office of Science, Office of High Energy Physics,  under contract No. DE-AC02-05CH11231.
ABS operates in the Parque Astron\'{o}mico
Atacama in northern Chile under the auspices of the
Comisi\'{o}n Nacional de Investigaci\'{o}n Cient\'{i}fica y
Tecnol\'{o}gica de Chile (CONICYT).
%
PWV measurements were provided by the Atacama
Pathfinder Experiment (APEX).
Some of the analyses were performed on the GPC supercomputer at the SciNet HPC
Consortium. SciNet is funded by the Canada Foundation of Innovation under the auspices of Compute
Canada, the Government of Ontario, the Ontario Research Fund --
Research Excellence; and the University of Toronto.
We would like to acknowledge the following for
their assistance in the instrument design, construction, operation, and
data analysis: G.~Atkinson, J.~Beall, F.~Beroz, S.~M.~Cho, B.~Dix,
T.~Evans, J.~Fowler, M.~Halpern, B.~Harrop,  M.~Hasselfield, 
J.~Hubmayr, T.~Marriage, J.~McMahon, M.~Niemack, S.~Pufu, M.~Uehara, and
K.~W.~Yoon.
We also thank very thorough reviewers for several
suggestions for improving the clarity of the paper.
T.~E.-H. was supported by a National Defense Science and Engineering Graduate Fellowship, as well as a National Science Foundation Astronomy and Astrophysics Postdoctoral Fellowship.
A.~K. acknowledges the Dicke Fellowship.
S.~M.~S. and K.~C. are supported by a NASA Office of the Chief Technologist's Space Technology Research Fellowship.
L.~P.~P. acknowledges the NASA Earth and Space Sciences Fellowship.

\bibliography{abs_hwp_sys}

\appendix

\section{Relation of Gauss-Hermite functions to beam-differencing experiments}
\label{sec:gh_to_diff}
We wish to convert between the Gauss-Hermite monopole, dipole, quadrupole basis used here and the differential gain, pointing, and ellipticity basis of Shimon et al. 2008.~\cite{2008PhRvD..77h3003S} The functions defined in Equations \ref{eqn:gh_low_order1}--\ref{eqn:gh_low_order3} can be related to the differential gain, differential beam width, differential pointing, and differential ellipticity for beam differencing experiments in the limit of small differences between the two beams. We note again that the differential beam width function, also a monopole term, is not induced by the HWP. An elliptical Gaussian offset from zero along the x axis can be denoted by
\begin{equation}
 G(\theta,\phi; g, \sigma, e, \rho) \equiv
  \frac{g}{2 \pi \sigma^2} \exp\left[ - \frac{\left(\theta \cos\phi - \rho
		     \right)^2}{2{\sigma^2 (1+e)}^2}
   - \frac{\left(\theta \sin\phi
		     \right)^2}{2{\sigma^2 (1-e)}^2}
      \right]  \:.
\label{eqn:ellip_gaussian}
\end{equation}
In the context of the differencing experiment, the intensity $I$ measurement
and linear-polarization $Q$ measurement are defined as
 $I = (T_x + T_y)/2$ and $Q = (T_x - T_y)/2$, where $T_x$ ($T_y$) denotes
 data from a detector sensitive to $x$ ($y$) polarization.
Two Gaussian beams with possible differences are associated to the measurements of $T_x$ and
$T_y$.  Thus, the template functions for the differential gain $D_{m0}$, differential width
$D_{m1}$, differential pointing $D_{d}$, and differential elipticity $D_{q}$ terms are
 \begin{eqnarray}
 D_{m0} & \equiv & 2 \pi \sigma^2 \cdot \frac{1}{2}
  \Bigl[G(\theta,\phi; 1+\frac{g}{2}, \sigma, 0, 0) \Bigr.  \nonumber \\
  & &\quad \quad \quad \quad \quad \quad \quad \quad - G(\theta,\phi; 1-\frac{g}{2}, \sigma, 0, 0) \Bigr] \nonumber \\
  & = & \Bigl. \frac{g}{2} \exp{ \left[- \theta^{2} / (2 \sigma^{2} )\right]}\:,
  \\
 D_{m1} & \equiv & 2 \pi \sigma^2 \cdot  \frac{1}{2} \Bigl[
  \frac{1}{(1+\mu)^{2}} G(\theta,\phi; \sigma(1+\mu), 0, 0) \Bigr.
  \nonumber
\label{eqn:diff_width_beam_subtraction}
  \\
 &&  \quad \quad \quad \quad \Bigl. - \frac{1}{(1-\mu)^{2}} G(\theta,\phi; \sigma(1-\mu), 0, 0) \Bigr]
  \:,
  \\
  D_{d} & \equiv & 
   2 \pi \sigma^2 \cdot \frac{1}{2} \Bigl[
   G(\theta,\phi; \sigma, 0, \rho/2) \Bigr. \nonumber \\
  & & \quad \quad \quad \quad \quad -
   G(\theta,\phi; \sigma, 0, -\rho/2) \Bigr] \:,
\label{eqn:dipole_beam_subtraction}
  \\
 D_{q} & \equiv & 2 \pi \sigma^2 \cdot  \frac{1}{2}
  \Bigl[ G(\theta,\phi; \sigma, e, 0) 
   -   G(\theta,\phi; \sigma, -e, 0) \Bigr] . \quad
\label{eqn:quad_beam_subtraction}
 \end{eqnarray}
Here, the prefactor $2 \pi \sigma^2$ comes from the fact that we fit the
beam maps (e.g., Figure~\ref{fig:maps}) that are normalized such that the
intensity $I$ beam peaks at unity;
it has the same origin as the $\sqrt{\pi\sigma^2}$ prefactor in
 Equation~(\ref{equ:def_monopole_quadrupole}).
The other dipole and quadrupole terms are derived from these by rotation by $90^{\circ}$ and $45^{\circ}$, respectively. 

For small pointing offsets, $(\rho / \sigma) \ll 1$, Equation \ref{eqn:ellip_gaussian} is approximately
\begin{equation}
 G(\theta,\phi; \sigma, 0, \rho) \simeq \left( 1 + \frac{\rho \theta \cos \phi}{\sigma^{2}} +  ... \right) \exp{\left(-\frac{\theta^{2}}{2 \sigma^{2}}\right)} .
\end{equation}

\noindent Putting this into Equation \ref{eqn:dipole_beam_subtraction} yields
\begin{equation}
D_{d} \simeq  \frac{ \rho \theta \cos \phi}{2 \sigma^{2}} \exp{\left(-\frac{\theta^{2}}{2 \sigma^{2}}\right)} .
\end{equation}

Similarly, a Gaussian with a small ellipticity $e \ll 1$ is approximately
\begin{equation}
 G(\theta,\phi; \sigma, e, 0) \simeq \left(1 + e \frac{\theta^{2} \cos 2 \phi}{\sigma^{2}} + ... \right)  \exp{\left(-\frac{\theta^{2}}{2 \sigma^{2}}\right)} .
\end{equation}

\noindent Substituting this into Equation \ref{eqn:quad_beam_subtraction} gives
\begin{equation}
D_{q} \simeq  \frac{e \theta^{2} \cos 2 \phi}{\sigma^{2}} \exp{\left(-\frac{\theta^{2}}{2 \sigma^{2}}\right)} .
\end{equation}

In calculating leakage coefficients, we are always taking ratios between the monopole and dipole or quadrupole terms. In Table \ref{tab:conversion_factors} we summarize the resulting functions including prefactors for the Gauss-Hermite basis versus the beam-subtraction definitions.

  \begin{table}[b]
  	\caption{ \label{tab:conversion_factors} %
    Summary of conversion factors between the leakage beam template functions
    defined by Gauss-Hermite functions and those of the beam subtraction formalism, where
    both have been normalized so that the main beam Gaussian has a maximum at unity. 
    Functions are defined in terms of the base Gaussian $G=\exp{\left[-\theta^{2}/(2 \sigma^{2})\right]}$.
    We only write out one dipole and one quadrupole function. The other is obtained by
    substituting sine for cosine. 
  	}
  	\begin{tabular}{p{7em}p{5em}p{6em}p{7em}}
  		\hline \hline
  		& Monopole & Dipole & Quadrupole \\
  		\hline
 Function &  $G$ &  $ ( \theta \cos \phi / \sigma) G $ & $ (\theta^{2} \cos 2 \phi/\sigma^{2}) G $ \\ \hline
 GH prefactor & $ a_{m0} $ & $ a_{d1} \sqrt{2}$ & $a_{q2}$  \\ \hline
 BS prefactor & $g/2$ & $\rho / 2 \sigma $ & $ e $ \\ \hline
 Conversion   & $a_{m0} = g/2$ & $a_{d} = \rho/\sqrt{8} \sigma $ & $a_{q} = e$ \\
  		\hline
  	\end{tabular}
  \end{table}

As opposed to the three functions discussed above, the
differential beam width and the function $f_{m1}(\theta,\phi)$
(Eq.~\ref{eqn:gh_low_order3}) do not have a one-by-one mapping.
A Gaussian with a small change in width is given by
\begin{equation}\begin{array}{cc}
 \frac{1}{1+\mu}G(\theta,\phi; \sigma (1 + \mu), e, 0)  \simeq  \\
	\hspace{1in}(1+\mu) \left( 1 + \mu\frac{\theta^{2}}{\sigma^{2}} \right) ( \exp{\left(-\frac{\theta^{2}}{2 \sigma^{2}}\right)}  \\
	\end{array}.
\end{equation}

\noindent Substitution into Equation \ref{eqn:diff_width_beam_subtraction} gives
\begin{equation}
D_{m1} \simeq  2 \mu \left( \frac{\theta^{2}}{\sigma^{2}} - 2 \right) \exp{\left(-\frac{\theta^{2}}{2 \sigma^{2}}\right)} .
\end{equation}
This corresponds to $2 \mu \left( f_{m1} - f_{m0} \right)$ in the
Gauss-Hermite formalism.

Thus, assuming that the monopole terms
 (those that do not vanish for $n=0$
in Eqs.~\ref{equ:radial_average_Q} and \ref{equ:radial_average_U})
can be expanded as a linear combination of $f_{m0}$ and $f_{m1}$,
or $D_{m0}$ and $D_{m1}$, we obtain the following relation between the
coefficients:
\begin{equation}
 \left(
  \begin{array}{c}
   a_{m0} \\
   a_{m1} \\
  \end{array}
	\right)
 =
 \left(
 \begin{array}{cc}
  1/2 & 2 \\
  0  & 2 \\
 \end{array}
 \right)
 \left(
  \begin{array}{c}
   g \\
   \mu \\
  \end{array}
	\right) \:.
\end{equation}
We recover the relation of $a_{m0} = g/2$ at the limit
of $a_{m1} \ll a_{m0}$.  This is true for the transfer-matrix model,
which predicts $a_{m1}=0$.
We also note that $f_{m0}$ and $f_{m1}$ are an appropriate basis set in
interpreting the results of Section~\ref{sec:scalar} since $f_{m1}$
integrates to
zero when integrated over $(\theta, \phi)$.

\end{document}